\documentclass[9pt]{article}

\usepackage[utf8]{inputenc}
\usepackage[T1]{fontenc}
\usepackage{lmodern}
\usepackage{float}
\usepackage{algorithm}
\usepackage{algpseudocode}

\usepackage{geometry}
\usepackage{graphicx}
\usepackage{amsmath, amssymb}
\usepackage{booktabs}
\usepackage{multirow}
\usepackage{caption}
\usepackage{xcolor}

\usepackage[numbers,sort&compress]{natbib}
\usepackage{xcolor}
\usepackage{hyperref}
\hypersetup{
    colorlinks=true,
    linkcolor=blue,
    citecolor=blue,
    urlcolor=blue
}

\newcommand{\edit}[1]{{\color{black}#1}}

\newenvironment{editViknesh}{\color{black}}{}

\usepackage{authblk}

\title{Wall Shear Stress Reconstruction from Concentration: Differentiable Physics and Physics-Informed Neural Networks}

\author[1,2]{Mahmoud Elhadidy}
\author[1,2]{Siva Viknesh}
\author[3]{Roshan M. D'Souza}
\author[1,2,4]{Amirhossein Arzani\thanks{Corresponding author: \texttt{amir.arzani@sci.utah.edu}}}

\affil[1]{Department of Mechanical Engineering, University of Utah, Salt Lake City, UT, USA}
\affil[2]{Scientific Computing and Imaging Institute, University of Utah, Salt Lake City, UT, USA}
\affil[3]{Department of Mechanical Engineering, University of Wisconsin--Milwaukee, Milwaukee, WI, USA}
\affil[4]{Department of Biomedical Engineering, University of Utah, Salt Lake City, UT, USA}

\date{}

\begin{document}

\maketitle

\begin{abstract}
Wall shear stress (WSS) governs near-wall transport dynamics and is a key hemodynamic indicator in cardiovascular flows, yet remains difficult to infer accurately due to the need for precise computation of near-wall velocity gradients. Passive scalar fields, such as concentration or temperature, are advected by the same underlying velocity field and have the potential to uncover hidden flow physics metrics such as WSS. In this work, we demonstrate such reconstruction from \edit{spatially limited passive scalar observations (confined to near-wall or far-field sub-regions of the domain)} using two fundamentally different inverse frameworks: a differentiable physics framework based on discrete adjoint, model-constrained optimization, which enforces the governing equations as \textit{hard} constraints, and physics-informed neural networks (PINNs), which treat them as \textit{soft} constraints. Benchmark problems include a two-dimensional (2D) canonical backward-facing step (2D-BFS) and a three-dimensional (3D) patient-specific stenotic coronary artery. For the 2D-BFS case, evaluated under three measurement scenarios (near-wall, far-field, and combined), PINN achieves high accuracy when near-wall data are available but fails when restricted to far-field measurements, whereas the differentiable physics approach recovers accurate WSS across all scenarios. In the 3D patient-specific case, the differentiable physics framework outperforms PINNs, yielding accurate WSS reconstruction. These results establish that measurement location and inverse formulation jointly determine reconstruction fidelity in scalar-based near-wall flow inference. The proposed framework opens a path toward estimation of near-wall hemodynamics from scalar transport data, with broader applicability to fluid flow problems where passive scalars can be observed.
\end{abstract}


% --------------------------------------------------

\section{Introduction}
Fluid flow near solid boundaries exerts tangential stresses that drive transport, structural response, and biological adaptation across a wide range of systems. Wall shear stress (WSS), the tangential force per unit area acting on a bounding surface, is the primary mechanical quantity connecting near-wall flow physics to these processes. In cardiovascular flows, WSS is especially important because it connects the endothelial cells, blood flow, and vascular health~\cite{tarbell2010shear,conway2013flow}. Through this mechanical stimulus, WSS induces mechanotransduction pathways, gene expression, and vascular remodeling~\cite{chatterjee2018endothelial}. Abnormal WSS patterns, such as low, oscillatory, or multidirectional shear, have been linked to the initiation and progression of atherosclerosis and aneurysms~\cite{meng2014high,morel2021effects,de2024predicting}. WSS also plays an important role in near-wall mass transport by affecting the exchange of oxygen, lipids, and other biochemical species between the blood and the vessel wall~\cite{mahmoudi2021story}. The importance of WSS is not limited to cardiovascular systems~\cite{schlichting1961boundary}. In aerodynamics, WSS determines skin-friction drag acting on aircraft and vehicles. In industrial pipe flows, it affects pressure losses and pumping power requirements. In environmental flows, WSS governs the onset of sediment transport and erosion in rivers and coastal regions.

Despite its importance, WSS is difficult to measure directly~\cite{katritsis2007wall,wang2020predicting}. Since WSS is proportional to the near-wall velocity gradient, its measurement requires resolving steep velocity variations within thin viscous boundary layers. This is experimentally challenging, particularly in biological and in-vivo settings~\cite{papaioannou2005vascular}.~\edit{While several techniques can estimate WSS, including particle image velocimetry, laser Doppler and hot-film anemometry, and shear-sensitive optical methods, they generally require optical or physical access, resolution of steep near-wall velocity gradients within thin boundary layers, or controlled experimental conditions, which makes accurate and direct WSS measurement particularly challenging in biological and in-vivo settings.} In contrast, passive scalars, such as dye concentration, temperature, or contrast-agent concentration, can often be measured with higher spatial resolution~\cite{crimaldi2008planar,wang2000high,woodworth2023heat,liao2023physics}. Because passive scalars are advected by the flow, their spatiotemporal evolution contains indirect information about the underlying velocity field. Therefore, scalar measurements can be leveraged to reconstruct flow quantities from transport data~\cite{sharma2019analytic,su1996scalar}. If the reconstructed velocity field is sufficiently accurate near the wall, WSS can then be obtained from the corresponding near-wall velocity gradients. This idea is also relevant to medical imaging, where contrast-agent transport in computed tomography angiography (CTA) and X-ray angiography provides an indirect marker of blood-flow hemodynamics. For example, contrast gradients in coronary CTA can encode flow-velocity information, while X-ray angiographic sequences can similarly exploit contrast propagation to estimate blood flow~\cite{lardo2015estimating,eslami2015computational}. Motivated by these observations, the present work considers the reconstruction of velocity fields, and subsequently WSS, from passive scalar transport measurements. Mathematically, the passive scalar evolution is governed by the advection--diffusion equation, and estimating velocity or boundary conditions from scalar observations leads to an inverse problem in which unknown flow quantities are inferred from available scalar measurements~\cite{raissi2020hidden}.

Several methods have been developed to solve such inverse problems. One common approach is adjoint-based optimization, where the governing equations are enforced as constraints and unknown parameters are estimated by minimizing the difference between simulated and observed data~\cite{gunzburger2002perspectives}. The adjoint equations need to be derived based on the problem of interest, gradients of objective functions with respect to quantities of interest calculated, and an iterative gradient-based optimization algorithm to be utilized to find the unknown quantities~\cite{giles2000introduction,kenway2019effective,mcnamara2004fluid}. Another approach is based on physics-informed neural networks (PINNs)~\cite{raissi2019physics}. In PINN, the solution fields are represented by implicit neural representation (neural fields) that are optimized to satisfy both the observational data and the governing physical equations. The governing equations are assembled in PINN using automatic differentiation and \edit{a gradient-based optimization approach, often using Adam} is employed. PINNs have shown strong potential for solving inverse problems and recovering hidden flow quantities from indirect observations~\cite{raissi2020hidden,arzani2021uncovering,jagtap2022physics,kim2024review,mai2024two}. A third group of approaches comes from data assimilation, including variational methods such as four-dimensional variational data assimilation (4D-VAR) and Kalman filtering variants, where model predictions and measurements are combined in a statistically consistent way to estimate the evolving state of the system~\cite{gaidzik2021hemodynamic,canuto2020ensemble,kalnay2003atmospheric,talagrand1987variational,evensen2009data,habibi2021integrating}. Finally, a direct learning approach could be utilized where data-driven neural networks or neural operators could be trained based on a large dataset to directly map measurements to quantities of interest such as WSS~\cite{elhadidy2026sle}.

Previous studies have explored the idea of inferring flow fields from passive scalar transport in different contexts. Early theoretical work analyzed the adjoint formulation of the advection--diffusion equation~\cite{buffoni2001adjoint}. Later, a space--time minimization approach was proposed for reconstructing velocity fields from scalar measurements~\cite{gillissen2018space}. Their scalar image velocimetry method formulates the reconstruction as an optimization problem constrained by the advection--diffusion equation and demonstrates that flow structures can be recovered even when scalar measurements are sparse or noisy. In biomedical imaging, similar ideas have been used to estimate blood flow from tracer transport data. Bakker et al.\ developed an advection--diffusion flow estimation method to infer coronary blood flow from contrast variations in CT angiography~\cite{bakker2021image}. Liu et al.\ proposed a framework for perfusion imaging, which estimates spatially varying velocity and diffusion fields by fitting an advection--diffusion model to time-series contrast data~\cite{liu2021perfusion}. Other work has reconstructed vascular velocity fields from tomographic projections using model-constrained optimization~\cite{huang2021reconstruction}. Experimental approaches such as ultrasound image velocimetry have also been used to estimate blood flow velocity and WSS by tracking speckle patterns generated by red blood cells in high-frame-rate ultrasound images~\cite{riemer2022contrast}. Similar ideas have also appeared in geophysical data assimilation, where tracer observations such as ozone concentrations can help constrain atmospheric wind fields~\cite{allen2018extraction}.

More recently, machine learning methods have been applied to these problems. Data-driven models have been used to reconstruct blood flow from concentration data~\cite{shusong2024deep,elhadidy2026sle}. PINNs have provided a physics-informed approach toward this mapping and the seminal Hidden Fluid Mechanics framework introduced by Raissi et al.\ showed that velocity and pressure fields can be reconstructed from passive scalar observations by enforcing the governing fluid flow equations during training~\cite{raissi2020hidden}. Subsequent studies applied PINN to several hemodynamic reconstruction problems, including estimation of velocity and pressure fields from magnetic resonance imaging (MRI) measurements~\cite{sarabian2022physics,sierpe2025estimation,toscano2025mr,kalajahi2025input} and reconstruction of coronary hemodynamics from angiography with uncertainty estimation~\cite{thakur2026punch}.

Although these studies show that flow quantities can be inferred from indirect measurements, most previous work assumes that observations are available throughout the entire spatial domain or that some velocity measurements are available to guide the reconstruction. In many experimental situations, however, measurements may only be available in a limited region due to imaging constraints, sensor placement, or measurement cost. When observations are confined to a small spatial window, the information available for reconstructing the velocity field becomes reduced, which can affect the accuracy of the inferred WSS. On the other hand, it may be postulated that given the dependence of WSS on near-wall flow structures~\cite{arzani2018wall}, a more efficient local approach focusing on the near-wall region could be leveraged for inverse modeling. Furthermore, despite the recent surge in PINN usage, a clear comparison between PINN and more traditional inverse modeling techniques (e.g., the adjoint method, which also employs equation-constrained optimization) has not been investigated in detail and the very few comparison studies focus on inferring full velocity from sparse velocity data~\cite{du2023state} rather than other surrogate measurements.

In this study, we investigate whether WSS can be reconstructed from passive scalar measurements that are available only in a limited region of the flow domain, without any direct velocity measurements, and without knowledge of inlet flow boundary conditions. We consider two fundamentally different inverse approaches: a discrete adjoint-based optimization framework implemented in FEniCS in a differentiable physics manner~\cite{mitusch2019dolfin}, and a PINN-based reconstruction method. By comparing these two approaches, we evaluate their reconstruction accuracy and robustness when the scalar observations are spatially restricted. We also discuss approaches for a local inverse modeling strategy focusing on the near-wall region. To the best of our knowledge, this work is the first to systematically study the reconstruction of WSS from passive scalar measurements confined to a limited spatial region and to provide a direct quantitative comparison between differentiable physics and PINN for inverse hemodynamic problems.

% --------------------------------------------------
\begin{editViknesh}

\section{Methods}
\label{sec:problem}

\subsection{Governing Equations}
\label{sec:problem-forward}

% $\Omega$, with boundary $\partial\Omega = \Gamma_{\mathrm{in}} \cup \Gamma_{\mathrm{out}} \cup \Gamma_{w}$, where $\Gamma_{\mathrm{in}}$, $\Gamma_{\mathrm{out}}$, and $\Gamma_{w}$ denote the inlet, outlet, and no-slip wall boundaries, respectively

Consider a steady incompressible Newtonian fluid flow through a wall-bounded domain. The velocity $\mathbf{u}$ and pressure $p$ fields satisfy the incompressible Navier--Stokes equations
\begin{align}
    \nabla \cdot \mathbf{u} &= 0\;,
    \label{eq:cont} \\
    \left(\mathbf{u} \cdot \nabla\right)\mathbf{u}
    &= -\frac{1}{\rho}\nabla p + \nu\,\nabla^{2}\mathbf{u} \;,
    \label{eq:mom}
\end{align}
where $\nu$ is the kinematic viscosity and $\rho$ is the density. In a forward problem, all boundary conditions are assumed known and typically include a prescribed velocity profile $ \mathbf{u} = \mathbf{u}_{\mathrm{in}}$ at the inlet, no-slip/no-penetration condition at the wall, and a prescribed outlet condition such as zero traction. 

% The boundary conditions are
% \begin{equation*}
%     \mathbf{u} = \mathbf{u}_{\mathrm{in}}
%     \text{ on } \Gamma_{\mathrm{in}} \;,
%     \qquad
%     \mathbf{u} = \mathbf{0}
%     \text{ on } \Gamma_{w} \;,
%     \qquad
%     \nu\,\frac{\partial \mathbf{u}}{\partial n} - p\,\mathbf{n} = \mathbf{0}
%     \text{ on } \Gamma_{\mathrm{out}} \;,
%     \label{eq:ns-bc}
% \end{equation*}
% where $\mathbf{n}$ is the outward unit normal. $\Gamma_{\mathrm{in}}$, $\Gamma_{\mathrm{out}}$, and $\Gamma_{w}$ denote the inlet, outlet, and no-slip wall boundaries, respectively. 

The steady velocity field $\mathbf{u}$ drives the transport of a passive scalar $C$, governed by the unsteady advection--diffusion equation 
\begin{equation}
    \frac{\partial C}{\partial t}
    + \mathbf{u} \cdot \nabla C
    = D\,\nabla^{2}C \;,
    \label{eq:adv-diff}
\end{equation}
where $D$ is the scalar diffusivity. The scalar $C$ represents a passive tracer,
such as a contrast agent or a temperature field, that is advected and diffused by the flow without altering the momentum balance. It should be noted that despite the steady nature of the momentum equation, the advection-diffusion equation is unsteady and represents the transient phase of injecting a passive scalar at the inlet of the domain $ C = C_{\mathrm{in}}$. Zero flux boundary condition is assumed for the tracer at the wall and outlet. The coupling between Eqs.~\eqref{eq:cont}--\eqref{eq:mom} and Eq.~\eqref{eq:adv-diff} is one-way: the velocity field $\mathbf{u}$ is first computed from the steady Navier--Stokes equations for a given inlet profile $\mathbf{u}_{\mathrm{in}}$, and is subsequently held fixed during the time integration of the scalar transport equation. 

%The complete forward map is written as

%\begin{equation}
%    C = \mathcal{F}(\mathbf{u}_{\mathrm{in}}),
%    \label{eq:fwd}
%\end{equation}
%where $\mathcal{F}$ represents the composition of these two solves.

The quantity of primary interest is WSS, which is evaluated as a post-processing step after the velocity field has been reconstructed. In the two-dimensional case, WSS (defined on the wall $\Gamma_{w}$) reduces to the following form

\begin{equation}
    \tau_{w}
    = \mu\,\frac{\partial u_{t}}{\partial n}
    \bigg|_{\Gamma_{w}},
    \label{eq:wss-2d}
\end{equation}
where $\mu$ is the dynamic viscosity, $u_{t}$ is the tangential component of the velocity, and $n$ denotes the wall-normal
direction.

In the three-dimensional case, the WSS is derived from the full stress tensor. The stress tensor is written as
\begin{equation}
    \boldsymbol{\sigma}
    = -p\,\mathbf{I}
    + \mu\left(
        \nabla\mathbf{u} + \left(\nabla\mathbf{u}\right)^{T}
    \right),
    \label{eq:cauchy}
\end{equation}
where $\mathbf{I}$ is the identity tensor. The traction vector acting on the wall is
\begin{equation}
    \mathbf{t} = \boldsymbol{\sigma}\,\mathbf{n}
    \big|_{\Gamma_{w}}\;.
    \label{eq:traction}
\end{equation}
The WSS vector is obtained by subtracting the normal component of the traction
\begin{equation}
    \boldsymbol{\tau}_{w}
    = \mathbf{t} - \left(\mathbf{t}\cdot\mathbf{n}\right)\mathbf{n} \;,
    \label{eq:wss-3d}
\end{equation}
such that $\boldsymbol{\tau}_{w}$ retains only the tangential stress contribution at the wall. Assuming known/constant viscosity~\cite{arzani2018accounting}, accurate WSS estimation requires accurate reconstruction of the velocity field and its gradient in the near-wall region.

\subsection{Inverse Problem}
\label{sec:problem-inverse}
Reconstructing WSS from scalar data is fundamentally challenging because WSS is determined by the near-wall velocity gradients, and these must be recovered indirectly from a scalar field that does not measure velocity directly. In practice, the inlet velocity profile $\mathbf{u}_{\mathrm{in}}$ is not known a priori, and neither velocity nor flow rate measurements are available anywhere in the domain. In this study, we assume that the only available data are spatially limited observations of the scalar concentration field, $ C^{\mathrm{data}}(\mathbf{x},t), \; \mathbf{x} \in \Omega_{\mathrm{obs}} $, where $\Omega_{\mathrm{obs}}$ is the observation region, which in general constitutes only a fraction of the full domain $\Omega$. 
The inverse problem is then to reconstruct either the velocity field $\mathbf{u}$ or the inlet velocity profile from $C^{\mathrm{data}}$ alone, and subsequently estimate wall shear stress $\boldsymbol{\tau}_{w}$, without any direct flow information.

Formally, this inverse problem is posed as the minimization of a loss functional that measures the mismatch between the predicted and observed scalar fields over $\Omega_{\mathrm{obs}}$, supplemented by a regularization term $\mathcal{R}$ that encodes prior knowledge and renders the otherwise ill-posed problem solvable.~\edit{The objective can, in general, be written using multiple temporal snapshots as follows, although the number of snapshots actually used differs by method:}

\begin{equation}
    \min_{\xi}\;
    \int_{0}^{T}\!\!\int_{\Omega_{\mathrm{obs}}}
    \left(
        C(\mathbf{x},t;\xi) - C^{\mathrm{data}}(\mathbf{x},t)
    \right)^{2}
    d\Omega\,dt
    \;+\;
    \int_{0}^{T}\!\!\int_{\Omega}
    \mathcal{R}\!\left(\xi,\mathbf{x},t\right)
    d\Omega\,dt \;,
    \label{eq:inverse_time}
\end{equation}
where $\xi$ denotes the optimization variable.~\edit{The regularization term $\mathcal{R}$ is written generically here because its concrete form depends on the specific inverse formulation adopted: each framework instantiates $\mathcal{R}$ differently according to its structure, as specified in the respective method sections.} In our problem, $\xi$ can correspond to the full velocity field or to a parameterized inlet velocity profile that drives the flow throughout the domain via the Navier--Stokes equations. Another choice is the manner in which the governing equations and boundary conditions are imposed, which could be soft or hard constraints. The difference between these choices distinguishes the two approaches considered in this work and is explained next.

% The general optimization structure of the scalar field reconstruction is summarized in Algorithm~\ref{alg:inverse}.
The two approaches investigated in this work (a differentiable physics solver and physics-informed neural networks) instantiate this procedure differently: they optimize different trainable parameters, enforce the governing equations and boundary conditions through distinct mechanisms, and employ different strategies to compute the gradients and execute the update step.

% \begin{algorithm}[H]
% \caption{Optimization algorithm for inferring WSS from scalar fields}
% \label{alg:inverse}
% \begin{algorithmic}[1]
% \State \textbf{Initialize} the method-specific trainable variables
% \Repeat
%     \State \textbf{Predict} the concentration field $C$
%            using the current flow estimate
%     \State \textbf{Evaluate} the loss by comparing $C$
%            against $C^{\mathrm{data}}$ on $\Omega_{\mathrm{obs}}$
%     \State \textbf{Update} the trainable variables to reduce the loss
% \Until{convergence}
% \State \textbf{Post-process:} extract WSS
%        $\boldsymbol{\tau}_{w}^{*}$ from the optimal reconstructed
%        velocity field $\mathbf{u}^{*}$
% \end{algorithmic}
% \end{algorithm}

%Our inverse problem is challenging because reconstructing wall shear stress requires recovering the near-wall velocity gradients, since WSS is determined by the viscous stress exerted by the fluid on the wall. Therefore, errors in the reconstructed velocity field, especially close to the wall, directly affect the accuracy of the WSS estimate. In this work, however, the velocity field is not measured directly and must be inferred from passive scalar observations, making the problem fundamentally ill-posed.

From the scalar advection--diffusion equation, Eq.~\ref{eq:adv-diff}, the velocity field enters scalar transport exclusively through the advection term $\mathbf{u}\cdot\nabla C$, as the projection of $\mathbf{u}$ onto the local concentration gradient $\nabla C$. The influence of the velocity field on scalar transport therefore depends on the alignment between $\mathbf{u}$ and $\nabla C$: it is maximum when they are aligned and vanishes when they are orthogonal, regardless of the velocity magnitude. The velocity field thus influences scalar transport more implicitly, through its projection onto $\nabla C$, rather than directly as a vector field. If the velocity moves fluid primarily along curves of nearly constant concentration, it produces little or no change in the observed scalar evolution. Hence, different velocity fields may produce similar scalar fields, especially when the scalar gradients are weak, smoothed by diffusion, or observed only in a limited region. This explains why scalar-only inversion is challenging and sometimes cannot be expected to recover the full velocity field without additional information~\cite{fiadeiro1984obtaining,wunsch1985can,sharma2019analytic}.

Although the Navier--Stokes equations and no-slip boundary condition reduce the set of physically admissible velocity fields, they do not completely remove the ill-posedness. Diffusion further compounds the difficulty by smoothing high-frequency scalar structures and irreversibly attenuating fine-scale advection signatures, making the inverse problem challenging. When $\Omega_{\mathrm{obs}}$ is located away from the near-wall region, the relationship between the observed scalar and the WSS becomes indirect and non-local. Thus, the reconstruction accuracy is expected to depend on the location of $\Omega_{\mathrm{obs}}$, which is a focus of the present study. 

%Lastly, it is imperative to remark that these arguments is specific to the ``one-way'' coupled setting considered here, in which the velocity field transports the scalar but is not influenced by it, and should be distinguished from two-way coupled formulations, such as those based on the Boussinesq approximation, where scalar variations can feed back onto the momentum equations.

% \subsection{Physics-constrained Data-driven Approaches for Inverse Modeling }
\subsection{Inverse Reconstruction Frameworks}
This section presents two inverse methodologies for inferring the WSS distribution from scalar concentration observations: PINN and differentiable physics solvers. Both approaches reconstruct the underlying velocity field through iterative optimization, but differ fundamentally in how the governing partial differential equations (PDEs) and known boundary conditions are enforced within the optimization loop. In PINN, both the governing PDEs and boundary conditions are imposed as \textit{soft constraints}, penalizing residuals of the governing equations through the loss functions. In contrast, the differentiable physics approach embeds a numerical PDE solver explicitly within the optimization loop, enforcing the governing equations and boundary conditions as \textit{hard constraints}, thus satisfying them up to the accuracy of the chosen discretization. A brief overview of the workflow of each method is shown in Fig.~\ref{fig:Methods_Comparison}.

%In the PINN formulation, the trainable variables correspond to the neural network parameters, and the velocity, pressure, and scalar fields are represented directly by neural network outputs as neural fields.  In the differentiable physics formulation, the trainable variables correspond to the inlet velocity profile, and the velocity and concentration fields are obtained by solving the governing equations. 

\begin{figure}[h!]
    \centering
    \includegraphics[width=0.6\textheight]{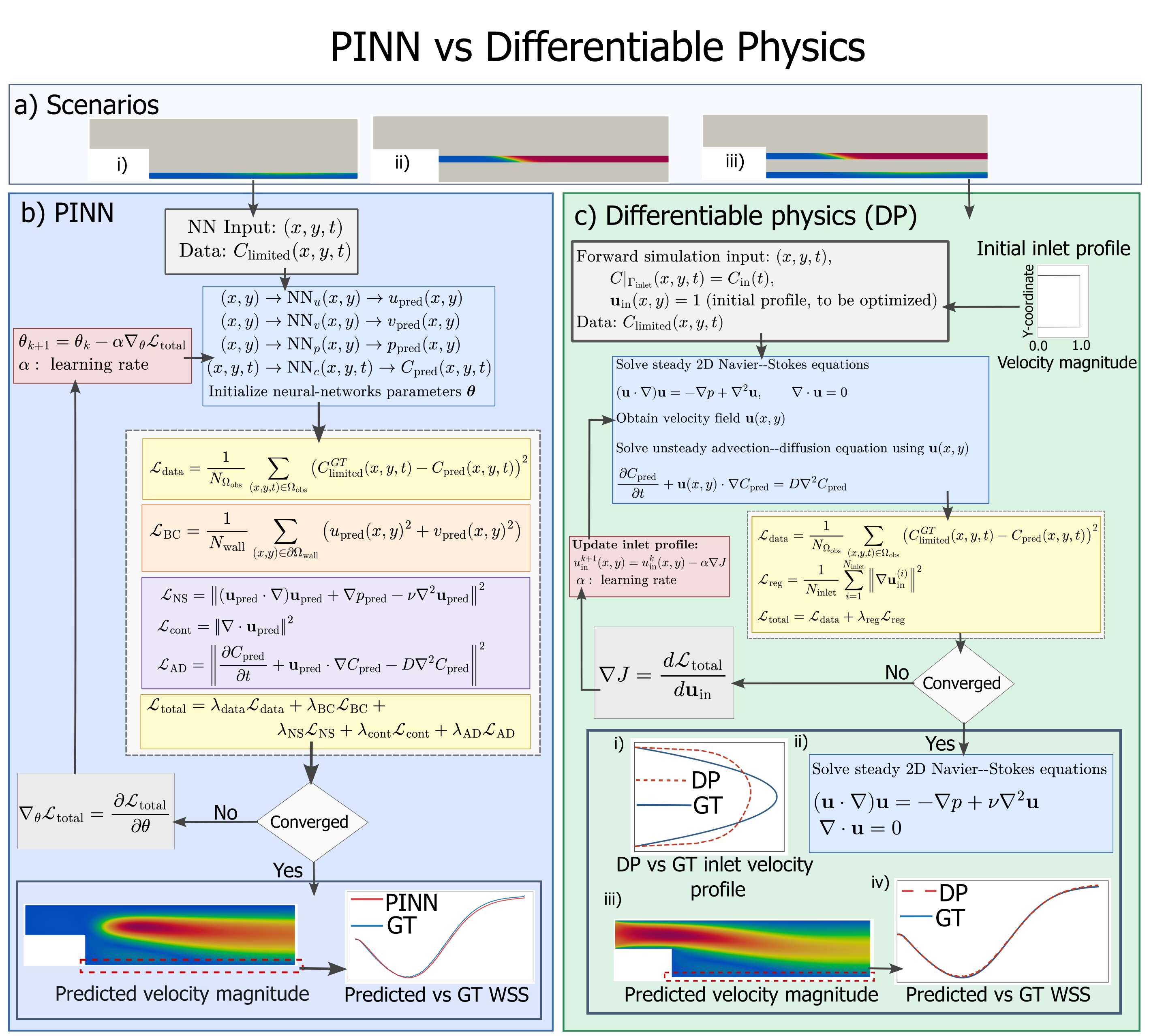}
    \caption{
    Comparison between physics-informed neural networks (PINNs) and differentiable physics (DP) for inverse reconstruction. 
    (a) Illustration of the three investigated concentration observation scenarios for the two-dimensional backward-facing step case. 
    (b) Overview of the PINN reconstruction framework, where the loss terms are constructed using the predicted fields and the limited concentration measurements. 
    Here, GT denotes ground truth. 
    (c) Overview of the DP approach, where the inlet velocity is optimized through the forward solver using the total loss gradient computed by the discrete adjoint and automatic differentiation.
    }
    \label{fig:Methods_Comparison}
\end{figure}

\end{editViknesh}
\subsubsection{Physics-Informed Neural Networks (PINNs)}

The PINN framework~\cite{raissi2019physics} is used to reconstruct the full velocity field from spatially confined unsteady concentration observations. The inverse problem is formulated as a direct reconstruction of the full velocity field from the available concentration data and the governing equations. In compact form, the goal is written as
\begin{equation}
    C_{\mathrm{data}}(\mathbf{x},t)\big|_{\Omega_{\mathrm{obs}}}
    \;\Longrightarrow\;
    \mathbf{u}_{\mathrm{NN}}(\mathbf{x})\big|_{\Omega} \;,
\end{equation}
where $\Omega_{\mathrm{obs}}$ is the observation region and $\Omega$ is the full domain.

The Navier--Stokes and advection-diffusion equations presented above are imposed as soft constraints through the physics loss. All required spatial and temporal derivatives are computed using backpropagation and automatic differentiation. The concentration data loss is evaluated only inside the observation region, where concentration data are available. Since the scalar measurements are unsteady, the data mismatch is integrated over both the observation region and the measurement time interval. In continuous form, this loss can be written as
\begin{equation}
    \mathcal{L}_{\mathrm{data}}
    =
    \int_{0}^{T}
    \int_{\Omega_{\mathrm{obs}}}
    \left|
    C_{\mathrm{NN}}(\mathbf{x},t)
    -
    C_{\mathrm{data}}(\mathbf{x},t)
    \right|^2
    \, d\mathbf{x}\, dt \;,
\end{equation}
where the integral is approximated by averaging the squared concentration error over the sampled spatial points in $\Omega_{\mathrm{obs}}$ and over all sampled time instances. In this study, the PINN data loss was evaluated using 20 unsteady
concentration snapshots for the 2D-BFS case and four unsteady concentration
snapshots for the 3D coronary artery case. We tested several numbers of
snapshots, and these values produced the best results for each case.

The physics loss is evaluated both inside the observation region and over additional points from the full domain. The contribution of the full-domain physics loss is controlled by the parameter $\alpha$
\begin{equation}
\label{eq:collocation_points}
    \mathcal{L}_{\mathrm{phys}}
    =
    \mathcal{L}_{\mathrm{phys}}^{\mathrm{in}}
    +
    \alpha \mathcal{L}_{\mathrm{phys}}^{\mathrm{out}} \;.
\end{equation}
When $\alpha=0$, only the physics residuals inside the observation region contribute to training. When $\alpha>0$, the model is also penalized for violating the governing equations outside the observed region.

The only velocity boundary condition imposed in the PINN formulation is the zero-velocity imposed at the wall. No concentration boundary conditions are prescribed in the final PINN formulation. In preliminary tests, we considered additional concentration boundary constraints, including a no-flux condition at the wall and a prescribed concentration value at the inlet. However, these constraints degraded the WSS reconstruction accuracy. Therefore, they were not included in the final objective function. A more detailed discussion of this behavior is provided in the Discussion section. The final training objective is written as
\begin{equation}
    \mathcal{L}_{\mathrm{total}}
    =
    \lambda_{\mathrm{data}}\mathcal{L}_{\mathrm{data}}
    +
    \lambda_{\mathrm{BC}}\mathcal{L}_{\mathrm{BC}}
    +
    \lambda_{\mathrm{phys}}
    \left(
    \mathcal{L}_{\mathrm{phys}}^{\mathrm{in}}
    +
    \alpha\mathcal{L}_{\mathrm{phys}}^{\mathrm{out}}
    \right) \;,
\end{equation}
where the different $\lambda$ parameters provide weights for different contributions. The neural network parameters are optimized using AdamW~\cite{loshchilov2017decoupled,kingma2014adam}. The detailed PINN architecture, sampling fractions, optimizer settings, learning-rate schedule, batch sizes, loss weights, and number of training epochs are reported in Appendix~\ref{AP:Hyperparameters}. The concentration data are only prescribed within the observation region 
$\Omega_{\mathrm{obs}}$; however, once trained, PINN defines a continuous 
approximation of the flow variables over the entire computational domain. 
This makes it possible to evaluate the reconstructed velocity field outside 
the measurement region and examine whether the physics-informed formulation 
can infer a physically consistent flow field in unobserved regions. 
The complete PINN reconstruction procedure is summarized in 
Algorithm~\ref{alg:pinn_reconstruction}.

\begin{algorithm}[h!]
\caption{PINN reconstruction from spatially confined scalar observations}
\label{alg:pinn_reconstruction}
\begin{algorithmic}[1]
\State \textbf{Initialize} the neural networks $C_{\mathrm{NN}}(\mathbf{x},t)$,
       $\mathbf{u}_{\mathrm{NN}}(\mathbf{x})$, and
       $p_{\mathrm{NN}}(\mathbf{x})$ weights
\State \textbf{Sample} scalar-observation points in
       $\Omega_{\mathrm{obs}}$, physics collocation points in
       $\Omega$, and boundary points on $\Gamma_{\mathrm{wall}}$
\Repeat
    \State \textbf{Predict} the concentration field
           $C_{\mathrm{NN}}(\mathbf{x},t)$, velocity field
           $\mathbf{u}_{\mathrm{NN}}(\mathbf{x})$, and pressure field
           $p_{\mathrm{NN}}(\mathbf{x})$
    \State \textbf{Evaluate} the scalar-data loss by comparing
           $C_{\mathrm{NN}}$ with $C_{\mathrm{data}}$
           in $\Omega_{\mathrm{obs}}$
    \State \textbf{Evaluate} the boundary-condition loss on
           $\Gamma_{\mathrm{wall}}$
    \State \textbf{Evaluate} the physics loss by penalizing the
           residuals of the governing equations
    \State \textbf{Evaluate} the gradient of the loss function 
           with respect to the parameters $\boldsymbol{\theta}$.
    \State \textbf{Update} the neural network parameters to reduce
           the weighted total loss
\Until{convergence}
\State \Return $C_{\mathrm{NN}}$, $\mathbf{u}_{\mathrm{NN}}$,
       and $p_{\mathrm{NN}}$
\State \textbf{Post-process:} compute WSS
       $\boldsymbol{\tau}_{w}^{*}$ from the reconstructed
       velocity field $\mathbf{u}_{\mathrm{NN}}^{*}$
\end{algorithmic}
\end{algorithm}
% In contrast, the DP approach optimizes only the inlet velocity profile. Once the inlet profile is specified, the velocity and pressure fields are generated by a forward Navier--Stokes solver and therefore satisfy incompressibility and momentum conservation as hard constraints. Optimizing the full velocity field directly within the DP framework would remove this advantage, since the resulting field would no longer be guaranteed to arise from a Navier--Stokes solution. Moreover, imposing the Navier--Stokes equations as additional soft constraints inside the DP optimization would be nontrivial in the present formulation because the pressure field is not explicitly optimized.

% The previous section described the traditional 3D full-order model (FOM) for solving this problem. We also explored a second approach that reconstructs WSS from passive scalar measurements using a surface transport model (STM) as a reduced model. Instead of embedding the full Navier--Stokes and advection--diffusion equations in the PINN loss, this approach uses a single surface transport equation introduced in~\cite{woodworth2023heat}. STM and a comparison with the FOM are presented in Appendix~\ref{app:stm}.

\subsubsection{Differentiable Physics (DP)}
%\subsubsubsection{Discrete Adjoint Framework}
The differentiable physics approach adopted in this work follows the differentiable programming paradigm~\cite{ramsundar2021differentiable,blondel2024elements}, in which the forward solver is embedded directly within the optimization loop such that gradients of the objective functional with respect to the optimization variables are computed automatically and remain strictly consistent with the discretized forward problem. In our work, this is achieved through a differentiable \textit{discrete} adjoint optimization framework built on the \texttt{dolfin} and \texttt{dolfin-adjoint} libraries~\cite{mitusch2019dolfin, farrell2013automated}, where the governing equations are solved using the finite element method (FEM) with the variational formulation expressed in the Unified Form Language (UFL), a high-level symbolic language, enabling the efficient computation of the sensitivity (gradient) of a quantity of interest with respect to a large number of design parameters, through the automated differentiation of the discretized PDE system.

Differentiability is facilitated by the \texttt{pyadjoint} backend, which operates by monitoring the forward solve and annotating each variational solve as it is executed, without being intrusive or requiring modifications to the forward model code. Each intercepted operation is recorded sequentially as a \texttt{Block} on the computational graph, referred to here as the global \texttt{Tape}, building a complete, ordered record of the forward computation. Each \texttt{Block} retains the critical information, such as the solution state and the variational form, required to construct the adjoint problem for that operation. The unknown field or parameter to be inferred is then registered as a \texttt{Control} variable. For example, in the present formulation, the inlet velocity profile is treated as the optimizable control, whereas quantities such as WSS are derived outputs computed from the reconstructed velocity field. The objective functional, which quantifies the misfit between the simulated and observed quantities, is then defined as a \texttt{ReducedFunctional}. This reduced functional represents the objective in terms of the control variable alone, with the state variables implicitly determined by the forward PDE solve. Thus, the optimization is carried out over the control variable, while the governing-equation constraints remain enforced through the forward model.

Gradient computation proceeds by replaying the recorded tape in reverse order to accumulate sensitivity information. For each recorded block, the framework applies symbolic differentiation to the underlying UFL variational form to assemble and solve the discrete adjoint system. The resulting adjoint variable is then passed to the preceding block, propagating gradient contributions via the discrete chain rule. A defining technical advantage of this method is that symbolic differentiation occurs \textit{prior} to discrete matrix assembly. This ensures the result is the exact adjoint of the assembled discrete PDE system, rather than a separately discretized version of a continuous adjoint equation. Consequently, the computed gradients are numerically exact and fully consistent with the discrete forward solver~\cite{farrell2013automated}.

%~\edit{In this sense, the DP framework is a discrete adjoint, PDE-constrained (variational) data-assimilation method. The comparison presented in this work is therefore directly a comparison between a discrete adjoint approach and PINN, the two frameworks differing primarily in whether the governing equations are enforced as hard or soft constraints.}

% Notably, a key distinction of this framework is the exact enforcement of the governing PDEs and prescribed (available) boundary conditions at the discrete level, ensuring satisfaction to machine precision. This contrasts with neural network–based approaches, such as PINN, in which the governing equations and boundary conditions are imposed implicitly through penalty terms in a loss functional, and exact satisfaction (i.e., perfect zero residuals) is generally unattainable. As a result, the present framework is particularly well-suited for engineering applications where strict adherence to conservation laws and boundary conditions is essential for solution physical validity. All computations are performed on CPU architectures, as GPU acceleration is not supported in this framework.

%\subsubsubsection{Inverse Formulation}

Within the differentiable physics framework, inferring the WSS distribution requires reconstructing the underlying velocity field from the measured scalar concentration field. Several strategies can be pursued, each with distinct trade-offs in physical fidelity and computational cost: $(i)$ In the first strategy, we solve only the unsteady advection--diffusion equation, treating the velocity at every nodal grid point in the domain as the optimization variable. While straightforward, this formulation is highly ill-posed and generally non-unique, because scalar observations constrain the velocity only through the advective term $\mathbf{u}\cdot\nabla C$. As a result, different velocity fields may produce similar concentration evolution~\cite{sharma2019analytic,raissi2020hidden}. Moreover, the recovered velocity field is not required to satisfy Navier--Stokes and continuity equations; $(ii)$ The second strategy attempts to address these deficiencies by adding the Navier--Stokes equations as soft constraints while again optimizing the velocity at all nodal grid points throughout the domain, and only solving the advection--diffusion equation in the forward solve. Although this improves physical consistency, these equations are imposed as soft penalties in the optimization objective rather than as hard constraints. Therefore, this strategy loses one of the main advantages of the differentiable physics formulation: exact satisfaction of the governing equations. In addition, evaluating the Navier--Stokes residual requires the pressure field, which is not directly observed and is not trivial to reconstruct.

For the present problem, we propose an alternative approach that is simultaneously physics-consistent, computationally efficient, and free of restrictive assumptions. Here, all governing equations are solved without compromise, but the optimization variable is restricted to the inlet velocity values at the boundary nodes, with no assumption imposed on the inlet profile shape. Since the continuity and Navier--Stokes equations are solved at every optimization step, mass and momentum conservation are automatically enforced, and the inlet condition uniquely governs the entire downstream flow field. The inlet velocity profile $\mathbf{u}_{\mathrm{in}}$ thus serves as the sole control variable, reducing the dimensionality of the inverse problem from the total number of domain nodes to the number of inlet nodes, while retaining full physical fidelity.

For the differentiable physics reconstructions, the data-misfit term was evaluated using only a single concentration snapshot, rather than a temporal sequence as used in the PINN framework. Specifically, the final time step was used for the 2D-BFS case, whereas an early snapshot at $t=0.05\,\mathrm{s}$ was used for the 3D coronary artery case to reduce computational cost. The inverse problem is formulated as the minimization of an objective functional comprising this single-snapshot data-misfit term and a regularization term

\begin{equation}
    J(\mathbf{u}_{\mathrm{in}})
    =
    \int_{\Omega_{\mathrm{obs}}}
    \left(
        C(\mathbf{x};\mathbf{u}_{\mathrm{in}}) 
        - C^{\mathrm{data}}(\mathbf{x})
    \right)^{2}
    \, d\Omega
    +
    \lambda
    \int_{\Gamma_{\mathrm{in}}}
    \left|
        \nabla_{\Gamma_{\mathrm{in}}} \mathbf{u}_{\mathrm{in}}(\mathbf{x})
    \right|^{2}
    \, d\Gamma \;,
\end{equation}
where $C(\mathbf{x};\mathbf{u}_{\mathrm{in}})$ is the concentration field predicted by the forward solver at location $\mathbf{x}$ and parameterized by the inlet velocity profile $\mathbf{u}_{\mathrm{in}}$, $C^{\mathrm{data}}(\mathbf{x})$ is the observed concentration field in $\Omega_{\mathrm{obs}}$, and $\lambda$ is a regularization parameter. The first term penalizes the misfit between the predicted and observed concentration fields within the observation region. The second term penalizes large spatial gradients (fluctuations) of the inlet profile along $\Gamma_{\mathrm{in}}$, suppressing non-physical oscillatory reconstructions. The gradient of the objective functional with respect to the control variable, $dJ/d\mathbf{u}_{\mathrm{in}}$, is computed via the discrete adjoint formulation described above, as implemented in \texttt{dolfin-adjoint}. The inlet velocity profile is subsequently updated using the Adam optimizer. The differentiable physics optimizer settings, learning rates, and regularization parameters used for each observation scenario are summarized in Appendix~\ref{AP:Hyperparameters}. The differentiable physics reconstruction procedure is summarized in Algorithm~\ref{alg:dp_reconstruction}. An ensemble optimization approach was adopted for the differentiable physics algorithm in each test case, which is discussed in Sections~\ref{subsec:2D-BFS} and~\ref{subsubsec:coronary_artery}.

% For the 2D-BFS case, a two-stage optimization strategy was employed. In the first optimization stage, the WSS reconstruction error remained between 18.2\% and 9.6\%, indicating that the solution could be trapped in a suboptimal local minimum. Therefore, in the second stage, the optimized inlet profile from each case was scaled and used to reinitialize the optimization. This procedure provided a better starting point and helped the optimizer move away from the initial local minimum, leading to improved WSS reconstruction.

%The DP reconstruction procedure is summarized in Algorithm~\ref{alg:dp_reconstruction}. In this framework, the inlet velocity profile $\mathbf{u}_{\mathrm{in}}$ is the control variable, while the velocity, pressure, and concentration fields are obtained from the forward Navier--Stokes and advection--diffusion solves. The gradient of the objective function with respect to the inlet profile is computed using the discrete adjoint method through \texttt{dolfin-adjoint}~\cite{mitusch2019dolfin,farrell2013automated}. The resulting gradient is then used by Adam to update the inlet profile. 

\begin{algorithm}[h!]
\caption{Differentiable physics reconstruction from spatially confined scalar observations}
\label{alg:dp_reconstruction}
\begin{algorithmic}[1]
\State \textbf{Initialize} the inlet velocity profile
       $\mathbf{u}_{\mathrm{in}}$ on $\Gamma_{\mathrm{in}}$
       and set it as the control variable
\Repeat
    \State \textbf{Solve} the governing equations using
           the current inlet profile $\mathbf{u}_{\mathrm{in}}$
           to obtain $\mathbf{u}$, $p$, and $C$
    \State \textbf{Evaluate} the scalar-data mismatch by comparing
           the predicted field $C$ with
           $C^{\mathrm{data}}$ in $\Omega_{\mathrm{obs}}$
    \State \textbf{Evaluate} the regularized total objective
           $J(\mathbf{u}_{\mathrm{in}})$
    \State \textbf{Compute} the gradient
           $dJ/d\mathbf{u}_{\mathrm{in}}$ using adjoint-based
           differentiation
    \State \textbf{Update} the inlet velocity profile
           $\mathbf{u}_{\mathrm{in}}$ to reduce the objective
\Until{convergence}
\State \Return $\mathbf{u}_{\mathrm{in}}^{*}$, $\mathbf{u}^{*}$,
       $p^{*}$ and $C^{*}$
\State \textbf{Post-process:} compute WSS
       $\boldsymbol{\tau}_{w}^{*}$ from the reconstructed
       velocity field $\mathbf{u}^{*}$
\end{algorithmic}
\end{algorithm}

\subsection{Benchmark Flow Problems}

\begin{figure}[h!]
    \centering
    \includegraphics[width=0.8\textwidth, height=0.85\textheight, keepaspectratio]{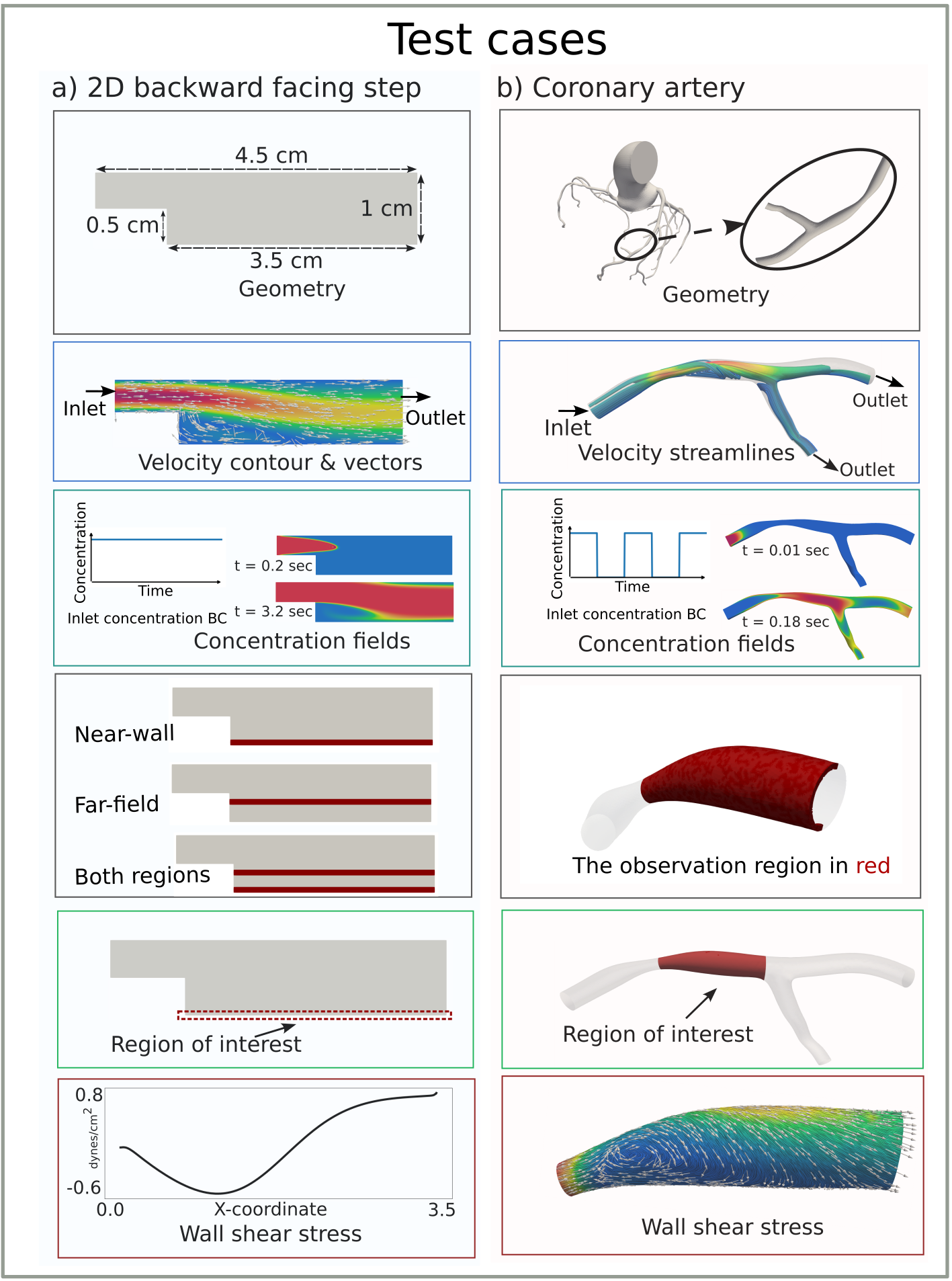}
    \caption{Overview of the test cases. (a) 2D backward-facing step (2D-BFS), showing geometry, velocity contours, concentration fields, three measurement scenarios, region of interest, and WSS. (b) Patient-specific coronary artery geometry, velocity streamlines, concentration fields, near-wall measurement region, region of interest, and WSS.}
    \label{fig:Test_Cases}
\end{figure}

\subsubsection{2D Backward-Facing Step}
\label{subsec:2D-BFS}

First, the flow through a classical two-dimensional (2D) Backward-Facing Step (2D-BFS) is considered. The geometric dimensions are as shown in Fig.~\ref{fig:Test_Cases}a. A parabolic velocity profile was prescribed at the inlet
\begin{equation}
u_{\mathrm{in}}(y) = U_{\max}\left(1 - \left(\frac{2y - H}{H}\right)^2\right),
\end{equation}
where \(U_{\max}=8~\mathrm{cm/s}\), \(H\) denotes the inlet channel height, and $y$ is the transverse coordinate measured across the inlet cross-section. Specifically $y=0$ corresponds to the lower inlet wall and $y=H/2$ is the channel centerline where the velocity reaches $U_{\max}$. With a mesh of $37,838$ triangular elements, the steady incompressible Navier--Stokes equations were solved using FEniCS with its built-in nonlinear solver to generate the ground-truth data. Using the computed steady velocity field, we then solved the transient advection--diffusion equation for a passive scalar concentration where $D = 0.001~\text{cm}^2/\text{s}$ was the diffusion coefficient. The Reynolds and P\'eclet numbers calculated based on the mean inlet velocity, the step height, and selected fluid properties were \edit{$Re = 71$ and $Pe = 2667$}, respectively. A constant concentration condition $C = 1$ was prescribed at the inlet, and natural zero-flux conditions were applied on all other boundaries. The advection--diffusion equation was advanced with a time step of $1\times10^{-3}$~s up to a final time of $4$~s, corresponding to $4000$ time steps to generate the data. The geometry, velocity, and representative concentration results are shown in Fig.~\ref{fig:Test_Cases}a.

The spatially limited concentration data obtained from this simulation are then used to reconstruct either the full velocity field directly or, equivalently, the inlet velocity profile that produced it. To this end, we investigate three different spatially limited measurement configurations (Fig.~\ref{fig:Test_Cases}a):

\begin{enumerate}
    \item Concentration data available only near the bottom wall, in a region with 
    a height of $0.1~\text{cm}$ (near-wall). 
    \item Concentration data available above the wall, in the region between 
    $0.3~\text{cm}$ and $0.4~\text{cm}$ (far-field).
    \item Concentration data available in both regions.
\end{enumerate}

The parameter $\alpha$ in Eq.~\eqref{eq:collocation_points} was varied for the near-wall and combined observation cases, but not for the far-field case. Varying $\alpha$ in these two cases allows us to assess whether enforcing the physics loss only in the observed regions is sufficient, or whether additional physics enforcement in the unobserved parts of the domain improves the reconstruction. In contrast, for the far-field case, setting $\alpha=0$ would apply the physics loss only away from the wall and leave the near-wall region, where WSS is calculated, without physics enforcement. For this reason, only the $\alpha=1$ case was considered for the far-field scenario.
For the PINN formulation, different neural networks are used to map spatial coordinates (x,y) to each velocity component and the pressure. In the case of concentration, time was also considered as input. The coordinate inputs were first mapped using Fourier feature encoding, also known as random Fourier features (RFF)~\cite{tancik2020fourier,sallam2023use,rahimi2007random} before being passed to the neural networks, which helps the networks represent sharper spatial and temporal variations in the solution. The encoded coordinates are then passed through fully connected neural networks with swish activation function. The 2D-BFS-specific PINN hyperparameters, including the network size, Fourier-feature settings, sampling fractions, and loss weights, are listed in Appendix~\ref{AP:Hyperparameters}.
After training, WSS is computed from the reconstructed velocity field and compared with the ground truth solution.

For the differentiable physics formulation in the 2D-BFS case, the control variable is the inlet velocity profile defined along the inlet boundary, $\mathbf{u}_{\mathrm{in}}(y)=\left(u_{\mathrm{in}}(y),v_{\mathrm{in}}(y)\right)$.
During the optimization, the degrees of freedom at the intersection of the inlet and the walls are clamped to zero to enforce no-slip at the inlet corners.

To facilitate convergence, the 2D differentiable physics optimization was performed using a two-stage restart strategy. In the first stage, the inlet velocity profile was optimized for 150 iterations starting from a low-magnitude uniform initial guess, $u_{\mathrm{in}}=1~\mathrm{cm/s}$. After this initial optimization, the resulting inlet profile was integrated across the inlet boundary to obtain the predicted flow rate, $Q_{\mathrm{pred}}$. In the second stage, $Q_{\mathrm{pred}}$ was used to define several scaled restart cases. For each scaling factor $s$, a new target flow rate was defined as
\begin{equation}
    Q_{\mathrm{new}} = s Q_{\mathrm{pred}} .
\end{equation}
Each restart case was initialized using a plug (uniform) inlet velocity profile whose magnitude produced the corresponding flow rate $Q_{\mathrm{new}}$. These scaled restart runs were then optimized for 100 iterations each. The restart case that produced the lowest loss magnitude after these 100 iterations was selected as the best initialization. Starting from this selected restart case, the optimization was then continued for an additional 150 iterations to obtain the final differentiable physics reconstruction. Therefore, the full procedure consisted of 150 iterations for the initial optimization, 100 iterations for each scaled restart candidate, and 150 additional iterations for the selected best restart. This strategy allowed the optimizer to explore nearby inlet conditions and reduced the likelihood of convergence to a poor local minimum. The learning rates and regularization parameters used for the differentiable physics reconstructions are provided in Appendix~\ref{AP:Hyperparameters}. Finally, WSS was calculated on the downstream bottom wall indicated as the region of interest in Fig.~\ref{fig:Test_Cases}a.

\subsubsection{3D Coronary Artery}
\label{subsubsec:coronary_artery}

The second test case consists of a three-dimensional (3D) patient-specific stenotic coronary artery geometry, where the geometry was utilized in our prior work~\cite{farghadan2019combined,mirramezani2019reduced}. The volumetric mesh was generated using the SimVascular mesher~\cite{updegrove2017simvascular} generating a tetrahedral mesh with 688,865 elements, and subsequently imported into the FEniCS solver. The incompressible Navier--Stokes equations were solved using FEniCS with the incremental pressure-correction scheme to solve the unsteady Navier--Stokes equations until the solution reached the steady state. The desired inlet velocity was not imposed instantaneously; instead, it was ramped linearly from zero to $U_{\max}=53$~cm/s and then held fixed until settled toward a steady state. Traction-free boundary condition was applied at both outlets. The fluid properties were set to $\mu = 0.04~\mathrm{g/(cm\cdot s)}$ and $\rho = 1.06~\mathrm{g/cm^3}$, with a time step of $\Delta t = 10^{-3}$~s. Once the velocity field was fully developed and nearly steady, the final quasi-steady flow solution was treated as the steady advecting velocity flow field for the scalar transport evolution.

The unsteady three-dimensional advection--diffusion equation was solved with diffusion coefficient $D = 0.005~\text{cm}^2/\text{s}$. The inlet concentration boundary condition was prescribed as a \emph{pulsed} signal with a total period of 0.5 s. Within each cycle, the concentration was initially injected and then alternated between on and off states every 0.1 s as shown in Fig.~\ref{fig:Test_Cases}b. Zero-flux conditions were applied on all remaining boundaries. The transport equation was advanced with a time step of $1\times10^{-4}$~s up to a final time of $0.5$~s. The resulting Reynolds number is \(Re = 211\), while the corresponding Péclet number is \(Pe = 1590\). For this coronary artery case, a single masking scenario was investigated, where measurements were available only in the near-wall region. A sectional view of the masked region is shown in Fig.~\ref{fig:Test_Cases}b.

The PINN formulation in the coronary artery case was similar to the 2D-BFS case but extended to three spatial dimensions. The hyperparameters are reported in Appendix~\ref{AP:Hyperparameters}. 

For the differentiable physics formulation in the coronary artery case, the same differentiable physics idea used for the 2D-BFS case is extended to the coronary artery case, but the inlet profile is now defined over the 3D domain's inlet surface. The 3D inverse solver mirrors the forward run described in this section. Before the optimization begins, a one-time Navier--Stokes warm-start is performed to reach the steady state before the inlet-control updates start. At each optimization iteration, the solver advances the flow field by solving the advection--diffusion equation till $t=0.05\,\mathrm{s}$ (when the concentration front reaches the region of interest) with the pulsed inlet concentration and evaluates the masked concentration data loss over the cells lying entirely within the near-wall region, computes the adjoint gradient via \texttt{dolfin-adjoint}, and updates the inlet profile with Adam. The degrees of freedom at the intersection of the inlet and the wall are clamped to zero so that the no-slip condition is respected at the inlet edge throughout the optimization. \edit{We also investigated the sensitivity to the selected observation time in Appendix~\ref{app:sensitivity}.}

To reduce sensitivity to the initial guess and improve convergence, the 3D case is initialized using an ensemble of uniform inlet velocities 
\begin{equation}
    \mathbf{u}_{\mathrm{in},k}^{0}
    =
    U_{k} \, \mathbf{n}_{\mathrm{in}},
    \qquad
    U_{k} \in [10,40]~\mathrm{cm/s} \;,
\end{equation}
where $\mathbf{n}_{\mathrm{in}}$ is the inlet normal direction. Each ensemble member is first run for a screening phase, and the best initialization (defined as minimum total loss) is selected as
\begin{equation}
    k^{*}
    =
    \arg\min_{k}
    J_{k} \;.
\end{equation}
This case is then continued for a longer optimization with
\begin{equation}
    \mathbf{u}_{\mathrm{in}}^{0}
    =
    \mathbf{u}_{\mathrm{in},k^{*}} \;.
\end{equation}

Each ensemble member was run for 100 iterations for screening, and the best case was continued to 300 iterations. After optimization, the reconstructed inlet profile was used in the forward solver to generate the velocity and WSS fields over the region of interest shown in Fig.~\ref{fig:Test_Cases}b. \edit{Mesh-independence studies for CFD simulation and WSS estimation are reported in Appendix~\ref{app:mesh}.}

%The key point in both the 2D and coronary artery DP cases is that we reconstruct the inlet velocity profile rather than the full velocity field. Once the inlet profile is known, the full velocity and pressure fields are generated by the Navier--Stokes solver, so the reconstructed flow field is physically consistent with the governing equations by construction.
% If we optimized the full velocity field directly, this guarantee would be lost. Even if the Navier--Stokes equations were added as soft constraints, the problem would remain difficult because the pressure field is not trivial to reconstruct and must be recovered consistently with the velocity field.

% --------------------------------------------------

\section{Results}
%In our study, we investigated two different methods to estimate WSS from passive scalar measurements without using any direct velocity or flow-rate measurements. Only concentration measurements within spatially confined observation regions were used. PINN reconstruct the full velocity field directly, while DP reconstructs the inlet velocity profile, which is then used in a full forward simulation to obtain the corresponding velocity field. In both cases, WSS is computed as a post-processing step.

For each of the two methods, we present qualitative and quantitative comparisons of the reconstructed WSS, which is the primary quantity of interest in this study. Velocity fields are also shown to provide physical context for the reconstructed flow physics. 
Quantitative errors are reported using the relative $L_2$ error, defined as
\begin{equation}
    \mathrm{Rel}\text{-}L_2 
    =
    \frac{
    \left\|
    \hat{\mathbf{y}} - \mathbf{y}
    \right\|_2
    }{
    \left\|
    \mathbf{y}
    \right\|_2
    },
    \label{eq:rel_l2_error}
\end{equation}
where $\hat{\mathbf{y}}$ denotes the predicted quantity and $\mathbf{y}$ denotes the corresponding ground-truth quantity.

\subsection{2D-BFS Results}
\label{sec:2D_Result}
For the 2D-BFS case, we investigated three scenarios. The first considers concentration measurements available only in the near-wall region. The second considers measurements available only in the far-field, advection-dominant region away from the wall. The third scenario combines both regions. In the differentiable physics approach, the full domain is used in the forward solver, while the data loss is enforced only within the specified measurement regions. In contrast, for PINN, we study the effect of using collocation points either within the measurement region only or across the rest of the domain. The contribution of this additional physics loss is controlled by the weighting parameter $\alpha$, as defined in Eq.~\eqref{eq:collocation_points}. Although velocity fields are shown for completeness, their accuracy is not the primary objective of this study. 
%In contrast, for DP, the optimized variable is the inlet velocity profile, from which the velocity field is obtained through a forward solve.

\begin{figure}[h!]
    \centering
    \includegraphics[width=\textwidth, height=0.85\textheight, keepaspectratio]{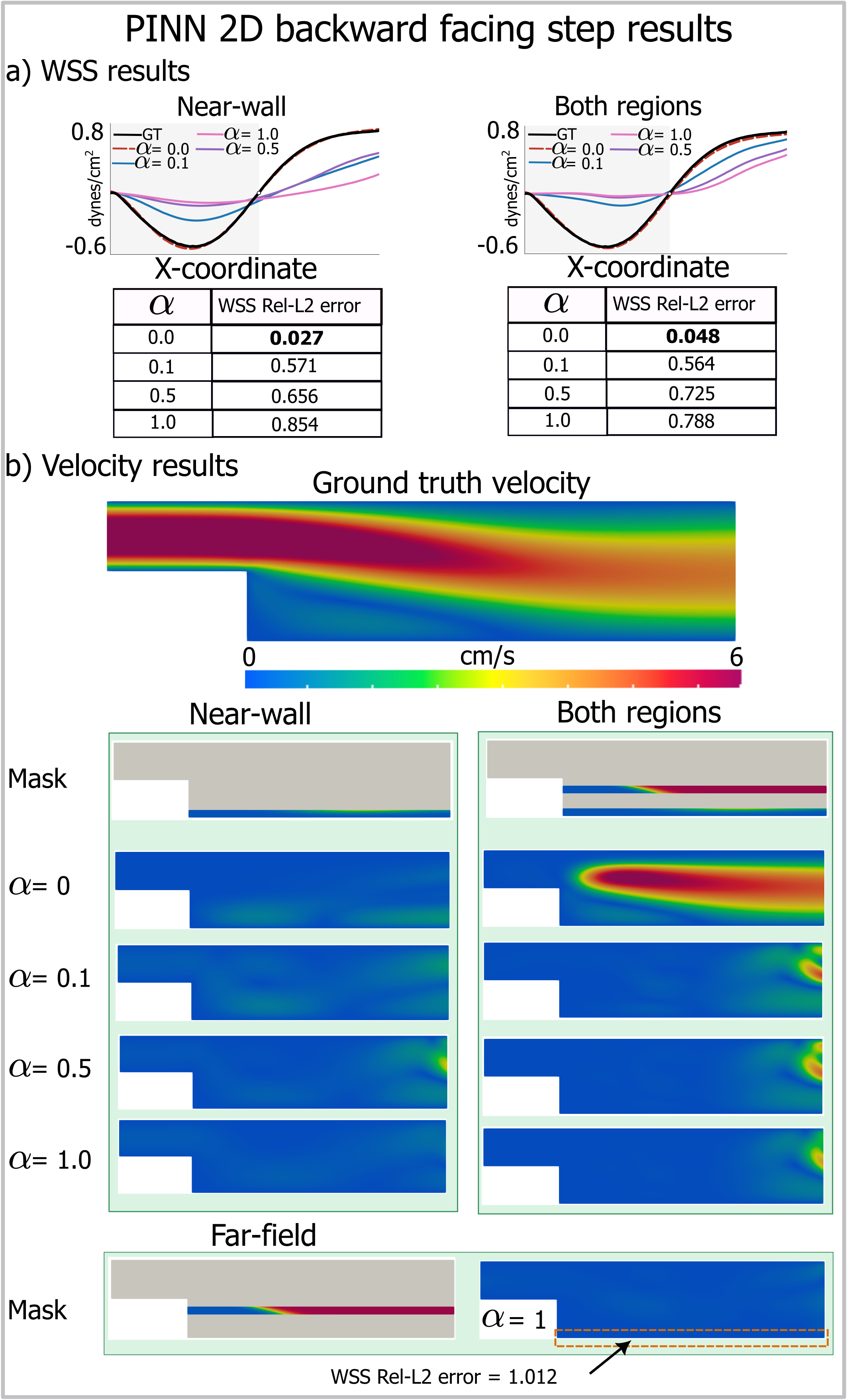}
    \caption{PINN results for the 2D-BFS case. (a) Comparison of predicted and ground-truth (GT) WSS. The tables report the Rel-$L_2$ error of the WSS for the corresponding values of $\alpha$. (b) Ground-truth velocity magnitude. Each column corresponds to a different observation scenario, while each row represents a different value of $\alpha$, which controls the weighting of physics-based collocation points outside the observation region.}
    \label{fig:2D-BFS_PINNs}
\end{figure}

\begin{figure}[h!]
    \centering
    \includegraphics[width=0.7\textwidth, height=\textheight, keepaspectratio]{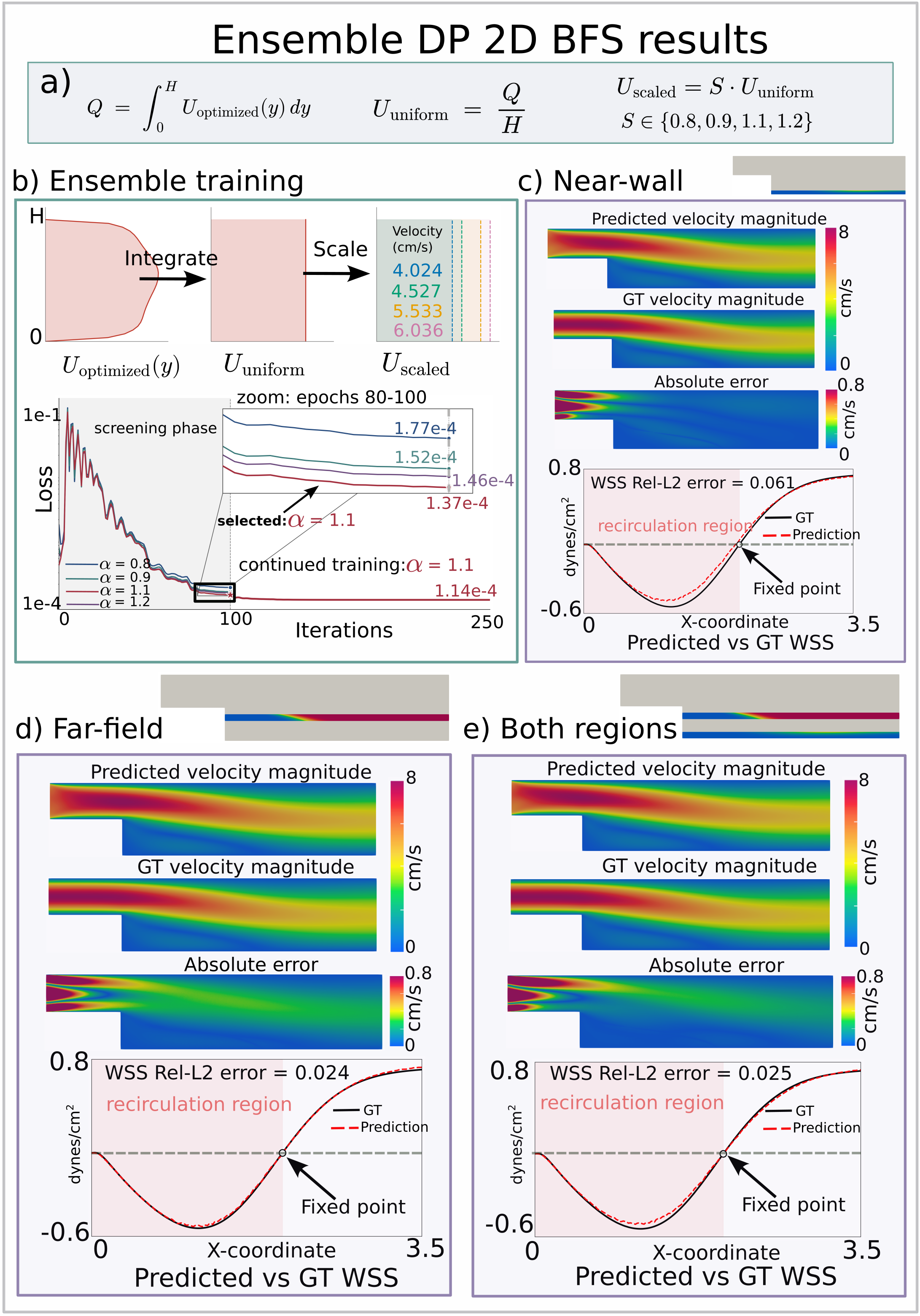}
    \caption{Ensemble differentiable physics results for the 2D-BFS case. 
    (a) Governing formulation and ensemble strategy. The inlet velocity is parameterized through the integrated flow rate obtained from a previously optimized profile. A uniform inlet velocity $U_{\mathrm{uniform}}$ is constructed to match the same flow rate, and multiple scaled variants $U_{\mathrm{scaled}}$ are generated to initialize the ensemble. Each initialization is evaluated during a screening phase based on the loss, after which the best-performing candidate is selected for further optimization. 
    (b) Ensemble training under near-wall observation shows the integrated flow rate and the loss evolution during the screening phase, followed by continuation of the run with the lowest loss. (c) The corresponding predicted velocity fields, WSS, and their comparison with ground truth. Panels (d) and (e) show the results for the far-region and combined-region observation cases, respectively, under the same ensemble training procedure.}
    \label{fig:ensemble_DP_2D_Results}
\end{figure}

For the near-wall case, PINN produced accurate WSS, with a Rel-$L_2$ error of $2.7\%$ in case of $\alpha$ = 0, as shown in Fig.~\ref{fig:2D-BFS_PINNs}a. However, increasing $\alpha$ increases the WSS error, indicating that adding collocation points from outside the observation region (which, here, coincides with the region of interest) negatively affects the reconstruction. Setting $\alpha = 0.1$ increases the error to more than $50\%$, reaching around $85\%$ when $\alpha = 1$.
Qualitatively, for $\alpha = 0$, there is almost perfect agreement between the ground-truth and predicted WSS. For higher values of $\alpha$, the magnitude becomes inaccurate, although most cases still capture the general behavior near the fixed point. For the velocity field (Fig.~\ref{fig:2D-BFS_PINNs}b), when $\alpha = 0$, the reconstruction is accurate near the wall of interest, but poor in the rest of the domain. As $\alpha$ increases, the velocity field becomes less physically consistent in the domain. For differentiable physics, the final ensemble-stage results are shown in Fig.~\ref{fig:ensemble_DP_2D_Results}. The WSS error is around $6\%$. Qualitatively, the WSS is underestimated in the recirculation region but becomes more accurate after the fixed point. For the velocity field, there is good agreement in the mainstream region, while the largest errors appear near the inlet. These results show that PINN achieves higher WSS accuracy when collocation points are restricted to the observation region as PINN can perform a fully local solution, while differentiable physics requires the full domain solution. 
%However, DP provides more accurate velocity reconstruction over most of the domain.

% \begin{figure}[h!]
%     \centering
%     \includegraphics[width=\textwidth, height=0.8\textheight, keepaspectratio]{DP_2DResults.png}
%     \caption{DP results for the 2D-BFS case. Panels (a)–(c) correspond to three different observation scenarios. In panel (a), where concentration measurements are available only near the wall, the ground-truth inlet velocity profile is compared with the predicted profile, along with the resulting fully developed velocity profiles. WSS distributions are also compared against the ground truth (GT). The bottom-right corner of each row reports the corresponding relative c$L_2$ errors.}
%     \label{fig:DP_2D_Results}
% \end{figure}

For the far-field case, $\alpha$ was not varied, as it is not reasonable to apply the physics loss only away from the wall where the data are located, while leaving the near-wall region, where WSS is calculated, without physics enforcement. Therefore, the only case shown is for $\alpha=1$. For this case, PINN fails to reconstruct WSS, with the Rel-$L_2$ error being around $100\%$.~\edit{This failure indicates that PINN cannot effectively leverage far-field, advection-dominated observations to constrain the near-wall velocity gradients that determine WSS, and the failure persisted under extensive tuning and experiments with second-order optimization.}
Qualitatively, the predicted WSS is almost flat. In contrast, differentiable physics achieves accurate results (Fig.~\ref{fig:ensemble_DP_2D_Results}d), with a low Rel-$L_2$ error of $2.4\%$ between the predicted and ground-truth WSS. The velocity results follow the same pattern as before, with the largest error near the inlet. Overall, this case shows that differentiable physics outperforms PINN, as PINN is not able to reconstruct accurate WSS or velocity fields from far-field concentration measurements alone.

In PINN, the combined regions scenario for $\alpha = 0$ has a WSS Rel-$L_2$ error of $4.8\%$. Although the predicted WSS distribution is qualitatively accurate, a small offset in magnitude contributes to the remaining error. The most notable improvement is observed in the downstream velocity reconstruction.~\edit{Since WSS is the target quantity, we focus on its reconstruction. As a secondary flow aspect, the downstream velocity field is better recovered when concentration data from both regions are used, although this is not the primary objective.}
Compared with the near-wall observation case, using concentration data from both regions helps PINN recover the downstream flow field more accurately. However, this improvement is mainly localized downstream of the step; the upstream velocity field is still not accurately recovered. Regarding WSS, increasing $\alpha$ leads to behavior similar to the near-wall case, with degraded accuracy. For differentiable physics, in the final results (Fig.~\ref{fig:ensemble_DP_2D_Results}e), the WSS is slightly underestimated, especially in the recirculation region, with final WSS reconstruction error of 2.5\%. The velocity results remain consistent with previous cases, with the largest errors near the inlet. Overall, final differentiable physics results are accurate and marginally lower than the far-field case. For PINN, this combined regions case reconstructed WSS with an error approximately twice that obtained in the near-wall case.

\subsection{3D Coronary Artery Results}
\label{sec:3D_Result}

\begin{figure}[h!]
    \centering
    \includegraphics[height=0.5\textheight, keepaspectratio]{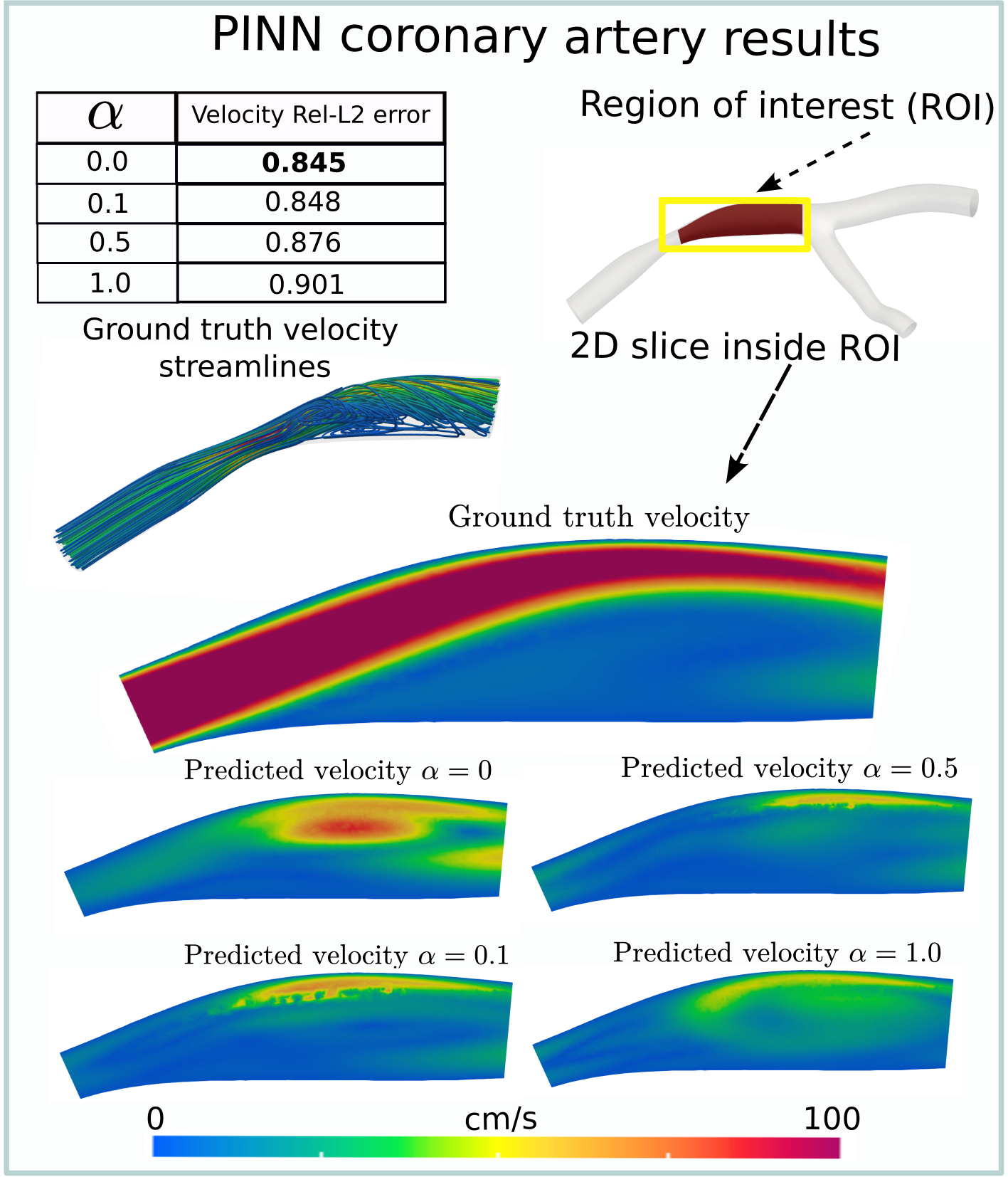}
    \caption{PINN results for the 3D coronary artery case. Comparison of the predicted and ground-truth (GT) velocity fields on a 2D slice within the region of interest. The table in the upper-left corner reports the Rel-$L_2$ error of the velocity field for different values of $\alpha$.}
    \label{fig:PINNs_LAD_VEL}
\end{figure}

\edit{For the 3D coronary artery case, the concentration measurements were confined to a single near-wall observation region located just downstream of the stenosis, as shown in Fig.~\ref{fig:Test_Cases}b.} Similar to the 2D-BFS case, PINN controlled the weight of the physics loss using the parameter $\alpha$, as defined in Eq.~\eqref{eq:collocation_points}. 
%For differentiable physics, the full domain was used in the forward solver, while the data loss was enforced only within the observation region. For the initialization of the inlet profile, an ensemble strategy was used: multiple runs were initialized with different uniform inlet velocity values, and only the run with the lowest loss was continued. For PINN, a similar approach to the 2D case was utilized.

\begin{figure}[h!]
    \centering
    \includegraphics[width=\textwidth, height=0.7\textheight, keepaspectratio]{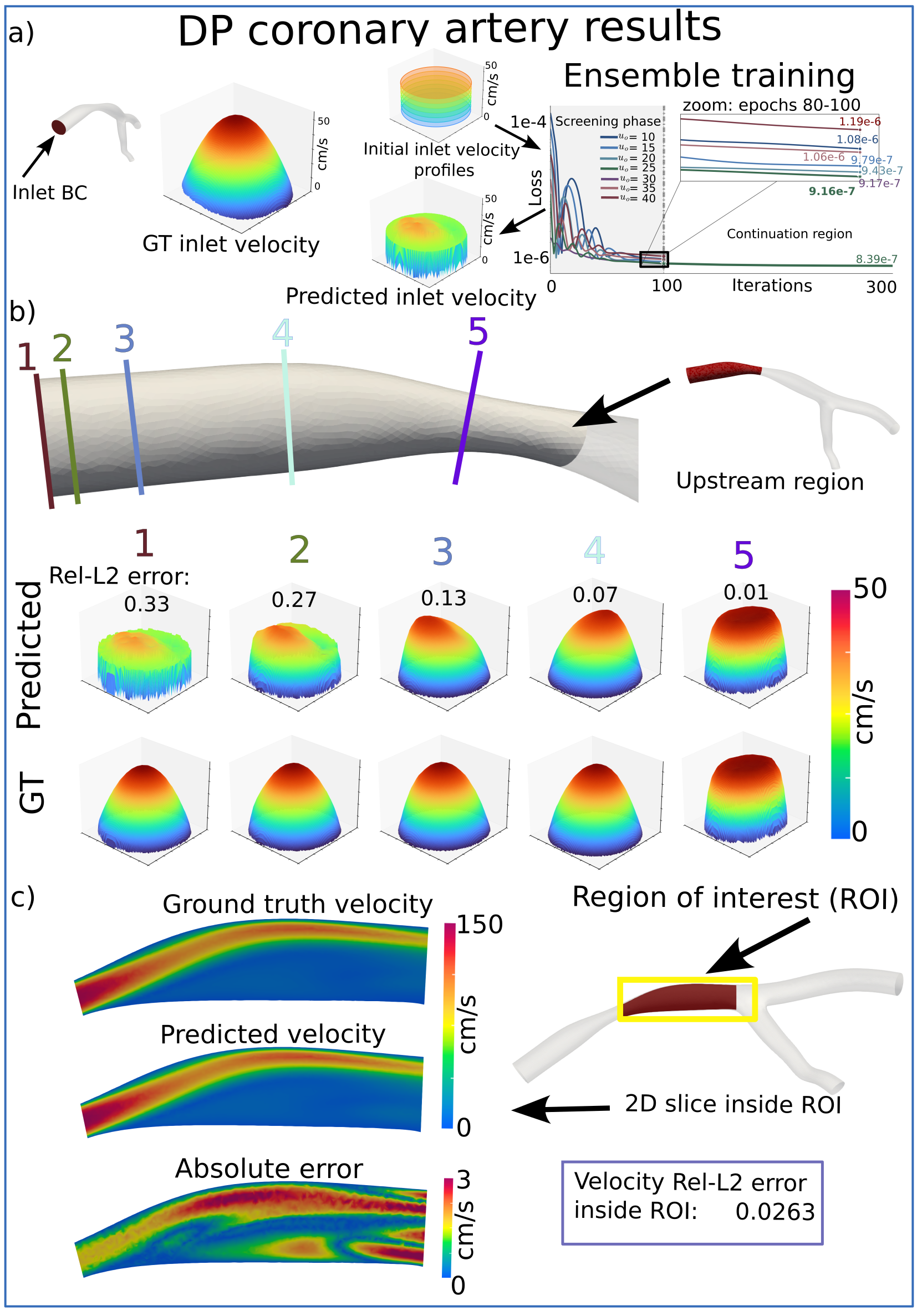}
    \caption{Coronary artery differentiable physics results. (a) Ensemble training procedure is summarized. The left panel shows the ground-truth (GT) inlet velocity profile. Seven different uniform inlet velocities were used as initializations, and all cases were optimized for 100 iterations. The initialization yielding the smallest loss was then continued to 300 iterations. The bottom-middle panel shows the final predicted inlet velocity profile. (b) Predicted inlet velocity profile development along the upstream region. The Rel-$L_2$ error for each slice is reported above the corresponding result. (c) Velocity field results at a slice located inside the region of interest (ROI) downstream of the stenosis. The Rel-$L_2$ error for the velocity field within the ROI was 0.0263.} 
    \label{fig:LAD_Adj}
\end{figure}

The velocity results for PINN are shown using a slice inside the region of interest in Fig.~\ref{fig:PINNs_LAD_VEL}. Since the observations are limited to the near-wall region, high velocity errors are expected away from the wall. The Rel-$L_2$ error, computed over the full 3D region of interest (not only the shown slice), ranges from $84\%$ to $90\%$, with the lowest error around $84\%$.

In contrast, differentiable physics achieves significantly more accurate velocity reconstruction, as shown in Fig.~\ref{fig:LAD_Adj}a. After optimization, the reconstructed inlet profile takes a plug-like shape. The evolution of this profile is examined along the vessel, from the inlet to the stenosis. While the inlet profile initially has a relatively high error of about $33\%$, it gradually develops into a much more accurate profile upstream of the stenosis, reaching an error of approximately $1\%$. This behavior is illustrated in Fig.~\ref{fig:LAD_Adj}b. As a result, the velocity error in the region of interest is reduced to around $2.6\%$, showing a clear improvement over PINN.

\begin{figure}[h!]
    \centering
    \includegraphics[width=0.7\textwidth, height=\textheight, keepaspectratio]{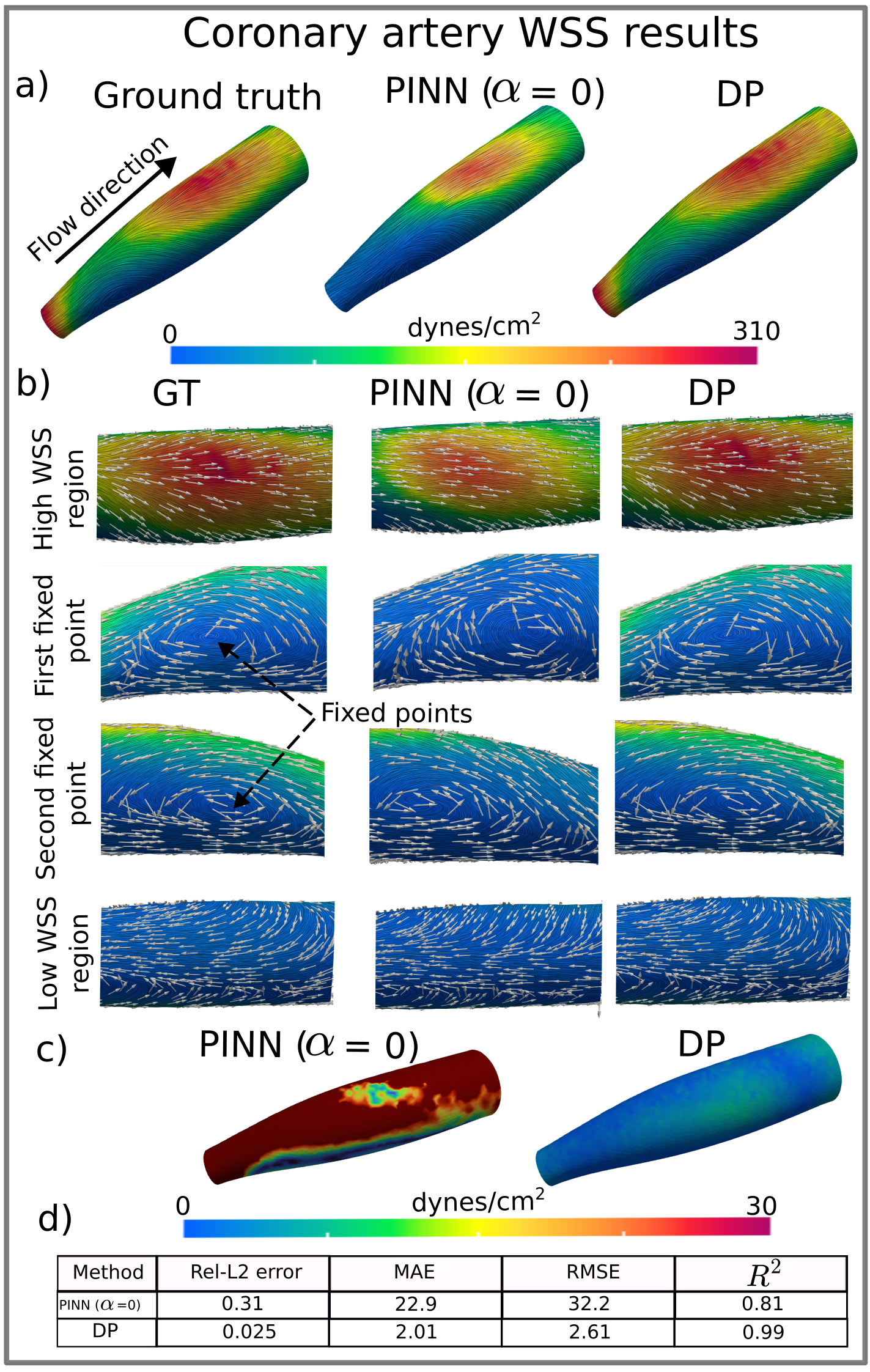}
    \caption{\edit{{Coronary artery WSS results. (a) Isometric views of the ground-truth (GT) WSS field, together with the PINN and differentiable physics predictions. The qualitative PINN results are shown for $\alpha = 0$. (b) Localized comparisons in representative regions of interest, including the high-WSS region, the first fixed point, the second fixed point, and the low-WSS region. (c) Spatial absolute error maps for the PINN and differentiable physics predictions, computed with respect to the GT WSS field. (d) Quantitative comparison using the coefficient of determination, defined as $R^2 = 1 - \sum_i (y_i - \hat{y}_i)^2 / \sum_i (y_i - \bar{y})^2$, where $y_i$ is the ground-truth WSS value, $\hat{y}_i$ is the predicted WSS value, and $\bar{y}$ is the mean ground-truth WSS value. The root mean squared error (RMSE) and mean absolute error (MAE) are also reported.}}}
    \label{fig:LAD_WSS}
\end{figure}

For WSS, which is the main quantity of interest, the qualitative comparison is shown in Fig.~\ref{fig:LAD_WSS}. From the isometric view, there is strong agreement in the main regions, especially between differentiable physics and the ground-truth. Both PINN and differentiable physics capture the overall WSS topology. However, differentiable physics shows better agreement in both magnitude and spatial distribution. A closer inspection of the WSS fixed-points~\cite{arzani2018wall} (second and third rows in Fig.~\ref{fig:LAD_WSS}) shows that PINN exhibits a slight shift in the location of the fixed points, while differentiable physics better preserves their positions. Additionally, while both methods show good alignment of WSS vector direction in most regions, PINN exhibits noticeable deviations in the low WSS region, particularly in the upstream part of the domain. Quantitatively, differentiable physics achieves an accurate WSS reconstruction, with a Rel-$L_2$ error of $2.5\%$. In contrast, PINN results in a significantly higher WSS error of $31\%$. 
%A similar trend is observed in the $R^2$ metric, where differentiable physics achieves excellent agreement with $R^2 = 0.99$, while the best PINN result reaches $R^2 = 0.81$.

% Overall, these results show that DP significantly outperforms PINN in the 3D patient-specific case. This improvement is mainly due to the ability of DP to reconstruct the inlet velocity profile and propagate the solution through a physically consistent forward solver. However, this comes at the cost of requiring the full geometry, a forward solve at each iteration, and concentration boundary conditions must be specified. In contrast, PINN directly reconstruct the solution from sparse observations, but they struggle to recover accurate velocity and underestimate WSS fields when only near-wall scalar measurements are available.

% --------------------------------------------------
\section{Discussion}

In this study, we compared two inverse formulations for reconstructing WSS from spatially limited passive scalar measurements, without using any velocity or flow-rate measurements. Two cases were considered: a benchmark 2D backward-facing step (2D-BFS) and a patient-specific stenotic coronary artery geometry. Both differentiable physics and PINN were applied to the same scalar measurements, but with different inverse formulations (hard-constraint formulation in differentiable physics versus soft-constraint formulation in PINN). \edit{The two frameworks are complementary (Appendix~\ref{app:guidance}, Table~\ref{tab:pinn_dp_guidance}). PINN needs no forward solver, full domain, or boundary conditions, suiting incomplete-geometry or local reconstructions, but is sensitive to measurement placement and hyperparameters and can fail when the data do not constrain the near-wall gradients. Differentiable physics enforces the equations exactly and is more accurate (especially in 3D) but needs the full geometry, concentration boundary conditions, and a forward/adjoint solve per iteration, making it costlier (Appendix~\ref{app:comp_cost}).}

For the 2D-BFS case, both methods achieved comparable WSS accuracy. PINN outperformed differentiable physics when near-wall measurements were used, and differentiable physics outperformed PINN in the far-field case. The differentiable physics required two optimization stages with an ensemble refinement stage, as shown in Fig.~\ref{fig:ensemble_DP_2D_Results}. Despite this additional cost, differentiable physics remained stable across all observation scenarios, whereas PINN failed when only far-field measurements were available. This behavior highlights the strong dependence of PINN-based reconstruction on the spatial location of the scalar measurements. The best observation region was different for each method. PINN performed best when near-wall data were available, because these measurements are more directly related to the near-wall velocity gradients that determine WSS. Differentiable physics achieved its best accuracy using far-field data, since the optimized variable is the inlet velocity profile and far-field advection-dominated measurements contain more direct information about the global flow. Interestingly, using the combined observation region did not always improve the reconstruction. Both methods showed a slight degradation compared with their best individual-region performance. This suggests that adding more scalar observations is not always beneficial when the added data come from regions with different sensitivities to the unknown flow field. In this case, near-wall and far-field measurements provide different types of information, and combining them made the problem harder to converge to the correct solution, likely due to the more challenging multi-task optimization.

Our investigation showed that the PINN reconstruction is sensitive to the fraction of concentration data used during training, the number of collocation points, and the inclusion of concentration boundary-condition constraints. In the 2D-BFS near-wall measurement case, the subsampled formulation without concentration boundary conditions produced the lowest WSS error of 2.7\%. When we tried to include concentration boundary conditions in the same subsampled case, including the wall no-flux condition and prescribed inlet concentration, the WSS error increased to 5.3\% (results not shown). Similarly, when we tried to use the full set of available concentration points instead of the subsampled training set, the WSS error increased to 5.5\% without concentration boundary conditions and to 6.2\% with concentration boundary conditions. Therefore, the main 2D-BFS PINN results consistently used the subsampled concentration points, with the corresponding sampling fraction reported in Appendix~\ref{AP:Hyperparameters}. One possible explanation is that the concentration boundary conditions do not add independent information for this inverse WSS reconstruction problem and make the optimization problem more challenging. In the near-wall observation case, the measured concentration field may already contain the effect of the wall no-flux condition. The inlet concentration condition is different because it is imposed upstream, while the available concentration measurements are restricted to a downstream near-wall region. Between the inlet and the observation region, there is a region with no concentration data. As a result, the inlet concentration boundary condition can only improve the reconstruction if the PINN accurately recovers the upstream velocity field and correctly transports the scalar from the inlet to the observed region. If this upstream velocity reconstruction is inaccurate, the inlet constraint provides limited benefit for the local WSS prediction and may instead introduce additional competing terms in the optimization. This interpretation may also explain why the subsampled case performed best. Subsampling can reduce the complexity of the optimization problem and focus the training on the most informative concentration points for WSS reconstruction. In contrast, using all concentration points can make the loss landscape more complex and may include many points that are less directly useful for the near-wall WSS target.
%Overall, these results suggest that the concentration boundary conditions are physically correct, but they do not necessarily improve the inverse WSS objective when their information is already encoded in the data or when the upstream transport needed to connect the boundary condition to the observation region is not accurately recovered.

In the coronary artery case, the difference between the two methods became more significant. PINN captured the overall WSS topology after extensive hyperparameter tuning, but the quantitative accuracy remained limited. Small offsets were observed in the fixed-point locations, and the WSS magnitude was underestimated. This indicates that recovering the qualitative WSS pattern is easier than recovering its exact magnitude. WSS magnitude depends strongly on accurate near-wall velocity gradients; therefore, small errors in the reconstructed velocity field can lead to larger errors in WSS. Differentiable physics clearly outperformed PINN in the coronary artery case. Since differentiable physics reconstructs only the inlet velocity profile and then solves the governing equations, the resulting velocity field satisfies the Navier--Stokes equations by construction, which leads to better agreement in WSS magnitude and topology. This advantage is particularly important in the coronary artery geometry, where the stenosis produces higher velocities, stronger gradients, and a more challenging inverse problem. However, differentiable physics also has practical limitations. It requires the full geometry, a full-domain forward solver, and imposing the correct concentration boundary conditions. The coronary artery results also showed sensitivity to the initial inlet profile, which motivated the ensemble initialization.
%Therefore, DP provides higher accuracy when the physical model and boundary information are available, while PINN provides a more flexible formulation when such information is incomplete, but with reduced quantitative accuracy in patient-specific cases.
% \edit{Conversely, DP fails when its assumptions are violated (incorrect geometry, boundary conditions). The methods also differ in minimum data: DP uses a single snapshot, since the forward model supplies the rest, whereas for PINN the snapshot count is a hyperparameter (20 in 2D, four in 3D).}

We also explored a second approach for reconstructing WSS from passive scalar measurements using a surface transport model (STM) as a reduced-order model within the PINN framework. This approach was motivated by the role of WSS vectors and WSS topology in near-wall transport~\cite{arzani2018wall,farghadan2019combined} and previously introduced surface transport models~\cite{arzani2016lagrangian,hansen2016reduced,chen2019relations}. Unlike the full-order model (FOM), which embeds the full Navier--Stokes and advection--diffusion equations in the PINN loss, the STM-based approach uses a single surface transport equation introduced in~\cite{chen2019relations,woodworth2023heat}. The STM formulation and its comparison with the FOM are presented in Appendix~\ref{app:stm}. This geometric reduced-order modeling approach reduces the computational cost, however, with a reduction in quantitative reconstruction accuracy as shown in Appendix~\ref{app:stm}.

Compared with previous PINN-based studies, such as Hidden Fluid Mechanics~\cite{raissi2020hidden}, our work considers a more restrictive and challenging setting. In particular, the available concentration measurements are spatially limited rather than provided over a broad observation domain. Moreover, the 3D coronary artery stenosis case involves localized high velocities, making the inverse reconstruction problem more difficult because advection-dominated transport produces sharper scalar gradients and stronger sensitivity to boundary and flow conditions. Our results are also consistent with their findings~\cite{raissi2020hidden}, where in the two-dimensional benchmark case, they achieved highly accurate velocity reconstruction, whereas the three-dimensional patient-specific case remained more challenging. Toscano et al.\ proposed magnetic-resonance artificial intelligence velocimetry (MR-AIV), a physics-informed framework to infer 3D velocity, pressure, and permeability from tracer concentration data~\cite{toscano2025mr}. Their results showed that MR-AIV recovered the main flow structures and spatial patterns, but accuracy decreased in more realistic and low-velocity regions, while absolute pressure and permeability magnitudes remained difficult to recover because of the ill-posed inverse problem. In addition, unlike studies that use sparse velocity measurements to reconstruct velocity and WSS fields~\cite{arzani2021uncovering,sarabian2022physics,du2023state,sierpe2025estimation}, our formulation relies only on passive scalar data.~\edit{The goal of a scalar-only formulation is not to outperform direct velocimetry when velocity is available, but to enable reconstruction when only a passive scalar can be observed.} This makes the inverse problem more challenging, because the scalar transport equation alone does not uniquely determine the full velocity field. An interesting aspect of the results is that the final WSS predictions are more accurate than the velocity field near the inlet, which aligns with the findings in~\cite{arzani2021uncovering}. Here, we show that the more important quantity is the inlet flow rate rather than the precise inlet velocity profile. Given the distance from the region of interest and the inlet, the flow reaches a relatively fully developed profile closer to the region of interest, which reduces sensitivity of the results on the precise uncovered inlet flow profile.~\edit{We emphasize this as a key finding. Because the region of interest is nearly fully developed, the near-wall gradients that set WSS depend mainly on the inlet flow rate, not the detailed inlet profile. This explains why WSS is reconstructed more accurately than the inlet profile itself, and mitigates the ill-posedness, since the target is insensitive to the non-unique components of the velocity field.}

%Sarabian et al.~\cite{sarabian2022physics} also demonstrated the use of PINN for brain hemodynamic reconstruction from sparse clinical measurements. Their framework combines limited transcranial Doppler velocity measurements with subject-specific vascular geometry from angiography and constrains the solution using a one-dimensional reduced-order model of pulsatile blood flow. They showed that the PINN approach can recover velocity, pressure, and vessel cross-sectional area across the cerebral vasculature despite limited measurement locations and uncertain inlet/outlet boundary conditions. Importantly, their results showed acceptable agreement with in vivo 4D-flow MRI validation data, while errors and deviations increased when relying on purely physics-based reduced-order simulations with uncertain boundary conditions. This supports the broader observation that physics-informed learning can improve inverse hemodynamic reconstruction from sparse data, but realistic patient-specific settings remain more challenging than simplified or well-constrained benchmark cases.

Several recent studies have addressed related inverse flow problems, but under stronger constraints than those used here. Domain-decomposed PINN for concentration-based velocity inference~\cite{ohashi2026physics} improves reconstruction by using larger datasets, extended domains, interface constraints, and prescribed flow rate boundary information. Similarly, physics-informed uncertainty-aware network for coronary hemodynamics (PUNCH)~\cite{thakur2026punch} reconstructs coronary flow from angiography by reducing contrast transport to a quasi-one-dimensional advection--diffusion problem along the vessel centerline. The method uses physiologically bounded neural network representations of axial velocity and effective dispersion, and incorporates uncertainty through a variational latent variable regularized by a Kullback--Leibler divergence term. Du et al.~\cite{du2023state} compared PINNs with 4D-VAR, an adjoint-based method, for reconstructing turbulent flow states from sparse velocity measurements. They showed that PINNs can recover the missing velocity field and infer the unobserved pressure field in turbulent channel flow; however, the PINN reconstructions were substantially less accurate than 4D-VAR, especially when predicting beyond the observation time window. Similarly, Sierpe et al.~\cite{sierpe2025estimation} showed that PINNs can estimate velocity, pressure, viscosity, and derived hemodynamic quantities
from realistic synthetic 4D-flow MRI measurements. However, their findings also indicate that reconstruction becomes more challenging for derivative-dependent quantities, such as WSS, particularly when observations are low-resolution, near-wall information is limited, or the flow state becomes more complex during diastole.

The observed sensitivity to measurement location is consistent with previous work showing that passive scalar-based reconstruction is strongly affected by the spatial distribution and density of the available measurements~\cite{rawden2026physics}. However, the setting in Rawden et al.~\cite{rawden2026physics} differs from the present study. Rawden et al. used PINNs to reconstruct mean velocity and passive scalar fields in a cylinder wake using both velocity and passive scalar measurements sampled on uniform grids with varied densities. In their cropped-domain source-identification cases, scalar data were supplied in a small source-centered subdomain, while velocity data were supplied over a larger surrounding subdomain. In contrast, our study reconstructs WSS from limited passive scalar observations alone, without sparse velocity measurements, inlet velocity data, or flow-rate constraints. Thus, although both studies show that measurement placement and data availability strongly influence reconstruction accuracy, the present inverse problem is more restrictive because the near-wall information is inferred only indirectly from scalar transport.

The advantage of the differentiable physics approach for inverse modeling is also supported by classical adjoint and data-assimilation theory. Adjoint formulations of the advection--diffusion equation show that the adjoint field propagates sensitivity information backward through the flow and identifies which upstream regions influence the observations~\cite{buffoni2001adjoint}. Similar ideas are used in 4D-VAR data assimilation, where tracer observations are used to update the underlying flow field through adjoint sensitivities~\cite{allen2018extraction,du2023state}.~\edit{In fact, our differentiable physics formulation can be viewed as a boundary-control variant of variational (4D-VAR-type) data assimilation: both minimize an observation-misfit functional with the governing PDEs enforced as hard constraints and gradients supplied by the adjoint, the difference being that we optimize the inlet control rather than an initial state. Our observation regarding the advantages of an adjoint approach are consistent with Du et al.~\cite{du2023state}, who report that PINN reconstructions are generally less accurate than adjoint/4D-VAR reconstructions.} 
%Our differentiable physics formulation uses automatic differentiation to assemble the matrices and record the nonlinear forward solver to compute the required gradients.
%Classical scalar image velocimetry also shows that scalar transport data alone only constrains the velocity component aligned with scalar gradients, making full velocity reconstruction intrinsically underdetermined without additional physical constraints~\cite{gillissen2018space,sharma2019analytic}.

%Overall, the results show that WSS reconstruction from spatially limited concentration measurements is strongly controlled by the inverse formulation and by the location of the measurements. PINN offer flexibility when boundary information is limited or the full domain is not accessible, but they struggle when the scalar data do not sufficiently constrain the velocity field. DP is less flexible and more computationally demanding, but it provides more accurate and physically consistent WSS reconstruction when the geometry, forward solver, and concentration boundary information are available. This distinction is especially important for patient-specific stenotic flows, where WSS depends on fine near-wall velocity gradients and is more sensitive than the velocity field itself.

% \edit{Several limitations of this study,}
Despite the promising results, several limitations should be noted. First, reconstructing velocity and WSS from passive scalar measurements is inherently ill-posed, since the advection--diffusion equation does not uniquely determine the underlying velocity field. As a result, multiple flow fields may produce similar scalar observations, particularly when the measurements are spatially limited.~\edit{Although the inverse problem for velocity is formally non-unique, reconstructions from different initializations and seeds converge to WSS fields with closely comparable errors (Appendix~\ref{app:uq}), indicating the target remains well constrained. Because the true WSS is unavailable in practice, reliability must be judged from observable quantities. We find that the concentration reconstruction error, measured against the available ground-truth concentration correlates with WSS accuracy, providing a practical, ground-truth-free confidence criterion (Appendix~\ref{app:conc_wss}).} In addition, the PINN-based formulation is sensitive to the temporal snapshots of the concentration field used during training. Different snapshot selections may contain different levels of information about the underlying flow, which can affect reconstruction accuracy. This introduces an additional source of variability that must be carefully considered in practical applications (selecting time-steps with the most information for quantities of interest). Also, several recent methodological advances in PINN~\cite{thawon2026physics,jnini2026curvature,wang2025simulating,costabal2024delta} were not investigated in the present study and will be considered in future work. The differentiable physics framework also introduces several practical constraints. It requires access to the full computational geometry, well-defined scalar transport boundary conditions (not needed in PINN), and knowledge of the physical time associated with each concentration measurement. In many real-world settings, such information may be incomplete or uncertain. Furthermore, the differentiable physics optimization is sensitive to initialization and may converge to local minima, which motivated the use of our proposed ensemble initialization strategy, which increases computational cost. The computational cost of differentiable physics can also become significant for large three-dimensional geometries, since each optimization step requires repeated forward and discrete adjoint solve with most of the available framework operated on CPUs. Another limitation is that the present results are obtained using clean, noise-free synthetic data,~\edit{although additional robustness experiments with noise and sparsity and an initial in vitro validation are provided in the Appendix (Appendix~\ref{app:noise_sparse}, Appendix~\ref{app:sensitivity} and Appendix~\ref{app:experimental}).} Finally, the velocity field was assumed to be steady in our main two test cases.

%\edit{The practical requirements differ between the two frameworks. Differentiable physics requires the full geometry, well-defined scalar boundary conditions, and the physical time of each measurement, whereas PINN avoids these but is more sensitive to measurement placement and hyperparameters. In clinical settings, these inputs are affected by segmentation error, uncertain boundary conditions and contrast administration, limited temporal resolution, and motion artifacts. To assess robustness, we evaluate both methods under sparse and noisy data, including combined noise and sparsity (Appendix~\ref{app:noise_sparse}), and analyze sensitivity to the observation region size and to an incorrect scalar diffusivity (Appendix~\ref{app:sensitivity}).}
%These limitations highlight the importance of measurement design, data quality, uncertainty quantification, and prior physical information when applying scalar-based flow reconstruction methods in realistic settings.

One possible future extension of the present study is dynamic computed tomography angiography (CTA) and X-ray angiography, where contrast-agent transport naturally provides time-resolved passive scalar information rather than direct velocity or WSS measurements. In dynamic CTA, the measured contrast evolution could be combined with patient-specific vascular geometry to infer hidden hemodynamic quantities, extending contrast-based flow estimation ideas toward local WSS reconstruction from limited scalar observations~\cite{lardo2015estimating,eslami2015computational,bakker2021image}. Similarly, in X-ray angiography, projection-based contrast intensity sequences could provide sparse observations of scalar transport, making our formulation relevant for estimating clinically important flow-related quantities without pressure wires or direct velocity measurements. This is consistent with recent angiography-based data-driven work for coronary microvascular dysfunction assessment, where contrast intensity profiles are used as informative surrogates for coronary physiology~\cite{yang2026assessing}. \edit{We emphasize that the present study is a controlled methodological comparison on synthetic benchmarks and, at present, not a deployable clinical tool. Therefore, claims of practical relevance should be read as motivation rather than demonstrated clinical utility. Nevertheless, to demonstrate potential for future translation, we demonstrate an in vitro validation case as reported in Appendix~\ref{app:experimental}.}

One promising direction for future work is the development of hybrid approaches that combine the flexibility of PINN with the physical consistency of differentiable physics. In such a framework, PINN can be used to provide an informed initial guess for the velocity field or inlet profile, which can then be refined using differentiable physics-based optimization, potentially improving convergence and reducing sensitivity to initialization or the need for an ensemble approach. Another important direction is the systematic study of measurement noise, data sparsity, and different concentration injection protocols. In practical applications, scalar observations are often noisy, temporally limited, or spatially incomplete. Quantifying the robustness of both PINN and differentiable physics under varying noise levels, snapshot selections, and sampling strategies will be essential for real-world deployment. This also includes investigating optimal measurement placement and identifying which regions of the flow domain provide the most informative measurements for WSS reconstruction. Designing optimal passive scalar injection strategies to gain most information for the inverse problem is an interesting problem for future work. In addition, incorporating prior information, such as physiological flow-rate constraints or auxiliary reduced-order models, may help mitigate the inherent ill-posedness of scalar-based reconstruction.~\edit{Another direction would be providing a rigorous analysis of the degree of non-uniqueness of the problem and the associated optimality conditions.}

% --------------------------------------------------
\section{Conclusion}
\edit{This study shows that WSS can be reconstructed from spatially limited passive scalar data, with accuracy depending on measurement location and the inverse formulation. These demonstrations use synthetic (CFD-based) ground truth together with initial in vitro validation, and full real-world clinical feasibility remains to be established.} In the 2D-BFS case, PINN performed well with near-wall data but failed with far-field measurements, while differentiable physics remained more robust across scenarios. In the 3D coronary case, differentiable physics outperformed PINN, producing more accurate velocity and WSS reconstruction under limited observations. Overall, differentiable physics provides stronger physical consistency but requires fuller domain knowledge, whereas PINNs are more flexible but more sensitive to data placement and training choices and hyperparameters, motivating future hybrid approaches.

\section*{Data Availability}

\edit{The implementation details and datasets used for the 2D backward-facing step (2D-BFS) and coronary artery cases are publicly available in the following GitHub repository:
\url{https://github.com/amir-cardiolab/DIfferentiable_Physics_wss}.
}

\section*{Conflict of Interest}
The authors declare no conflicts.

\section*{Acknowledgments}
This research was funded by the National Science Foundation (NSF) Award No. 2205265.

\appendix
\section{Appendix}

\subsection{Hyperparameters}
\label{AP:Hyperparameters}

The PINN hyperparameters used for the 2D-BFS and the 3D coronary artery cases are summarized in Table~\ref{tab:pinn_hyperparameters}. In both cases, AdamW was used with a step learning-rate scheduler, and the loss combined data mismatch, no-slip wall boundary conditions, and physics residuals from the incompressible Navier--Stokes and scalar transport equations.

\begin{table}[h!]
\centering
\caption{PINN hyperparameters used for the 2D-BFS and coronary artery cases.}
\label{tab:pinn_hyperparameters}
\begin{tabular}{lcc}
\hline
\textbf{Hyperparameter} & \textbf{2D-BFS} & \textbf{Coronary artery} \\
\hline
Hidden neurons & $100$ & $70$ \\
Hidden layers & $8$ & $8$ \\
Fourier frequencies & $10$ & $10$ \\
Fourier feature standard deviation & $1.0$ & $0.1$ \\
Activation function & Adaptive Swish & Swish \\
Optimizer & AdamW & AdamW \\
Initial learning rate & $10^{-3}$ & $10^{-3}$ \\
Epochs & $3000$ & $6000$ \\
Data batch size & $1024$ & $1024 \times 6$ \\
Physics batch size & $1024 \times 8$ & $1024 \times 6$ \\
Data sampling fraction & $0.05$ & $1.0$ \\
Physics/collocation sampling fraction & $0.25$ & $1.0$ \\
Scheduler decay rate & $0.6$ & $0.5$ \\
Scheduler step size & $300$ & $1000$ \\
$\lambda_{\mathrm{data}}$ & $10^{5}$ & $2 \times 10^{5}$ \\
$\lambda_{\mathrm{BC}}$ & $1000$ & $100$ \\
$\lambda_{\mathrm{cont}}$ & $200$ & $25$ \\
$\lambda_{\mathrm{mom}}$ & $200$ & $10$ \\
$\lambda_{\mathrm{adv-diff}}$ & $200$ & $10$ \\
\hline
\end{tabular}
\end{table}

The differentiable physics inverse reconstruction was optimized directly over the inlet velocity profile as the control variable. The hyperparameters are mentioned in Table~\ref{tab:dp_hyperparameters}.

\begin{table}[h!]
\centering
\caption{Differentiable physics hyperparameters used for the 2D-BFS and coronary artery cases.}
\label{tab:dp_hyperparameters}
\resizebox{\textwidth}{!}{%
\begin{tabular}{lcccc}
\hline
\textbf{Hyperparameter} & \textbf{2D-BFS (Near-wall)} & \textbf{2D-BFS (Both regions)} & \textbf{2D-BFS (Far-field)} & \textbf{Coronary artery} \\
\hline
Optimizer & Adam & Adam & Adam & Adam \\
Initial learning rate (LR) & $2.0$ & $2.5$ & $2.5$ & $2.0$ \\
LR decay rate & $0.4$ & $0.4$ & $0.4$ & $0.5$ \\
LR decay steps & $8$ & $8$ & $8$ & $6$ \\
Regularization weight & $5.0 \times 10^{-6}$ & $2.0 \times 10^{-5}$ & $2.0 \times 10^{-5}$ & $1.0 \times 10^{-8}$ \\
\hline
\end{tabular}%
}
\end{table}

\subsection{Surface Transport Model (STM)}
\label{app:stm}

Typically, WSS is related to passive scalar measurements in two steps. First, an inverse problem is solved to reconstruct the velocity field from the measured scalar field. Then, WSS is computed as a post-processing step from the reconstructed near-wall velocity gradients. An alternative approach was introduced in~\cite{chen2019relations,woodworth2023heat}, where a surface transport equation is derived and used to relate the WSS directly to surface measurements. In this formulation, the advection--diffusion equation is projected onto the wall, and a near-wall Taylor-series expansion is used for the velocity and scalar fields.

Near a flat wall (x-z plane), the tangential velocity components can be approximated as
\begin{equation}
    u_x \approx \frac{\tau_x}{\mu}y \;,
    \qquad
    u_z \approx \frac{\tau_z}{\mu}y \;,
\end{equation}
where $y$ is the wall-normal direction, $\mu$ is the dynamic viscosity, and $\tau_x$ and $\tau_z$ are the WSS components. The scalar field $C$ is also expanded near the wall as
\begin{equation}
    C(x,y,z,t)
    =
    C(x,z,t)
    +
    \left(\frac{\partial C}{\partial y}\right)_w y
    +
    \frac{1}{2}
    \left(\frac{\partial^2 C}{\partial y^2}\right)_w y^2
    +
    \frac{1}{6}
    \left(\frac{\partial^3 C}{\partial y^3}\right)_w y^3
    + \cdots \;,
\end{equation}
where \(C(x,z,t)\) is the surface concentration. By substituting these near-wall expansions into the advection--diffusion equation and differentiating with respect to the wall-normal direction, the wall limit gives a direct relationship between the WSS components and the surface scalar gradient
\begin{equation}
    \tau_x \frac{\partial C}{\partial x}
    +
    \tau_z \frac{\partial C}{\partial z}
    =
    \mu D
    \left(
    \frac{\partial^3 C}{\partial y^3}
    \right)_w ,
\end{equation}
where $D$ is the scalar diffusivity. In compact notation, this can be written as
\begin{equation}
    \boldsymbol{\tau}_w \cdot \nabla_s C
    =
    \mu D
    \left(
    \frac{\partial^3 C}{\partial y^3}
    \right)_w ,
\end{equation}
where $\nabla_s C = \left(\partial C/\partial x,\partial C/\partial z\right)$ is the surface gradient of the scalar field. Therefore, instead of first reconstructing the full velocity field and then computing WSS, this approach directly links WSS to measurable surface scalar gradients and wall-normal scalar derivatives and represents a surface transport model (STM).
%~\edit{Because this derivation relies on a near-wall Taylor expansion about a locally flat wall, the STM in its present form is restricted to (locally) planar surfaces and cannot be applied directly to the curved coronary geometry. It is therefore presented as a separate reduced-order demonstration and benchmarked against a full-order PINN, rather than used as a baseline for the two main test cases.}

\begin{figure}[h!]
    \centering
    \includegraphics[width=\textwidth, height=0.74\textheight, keepaspectratio]{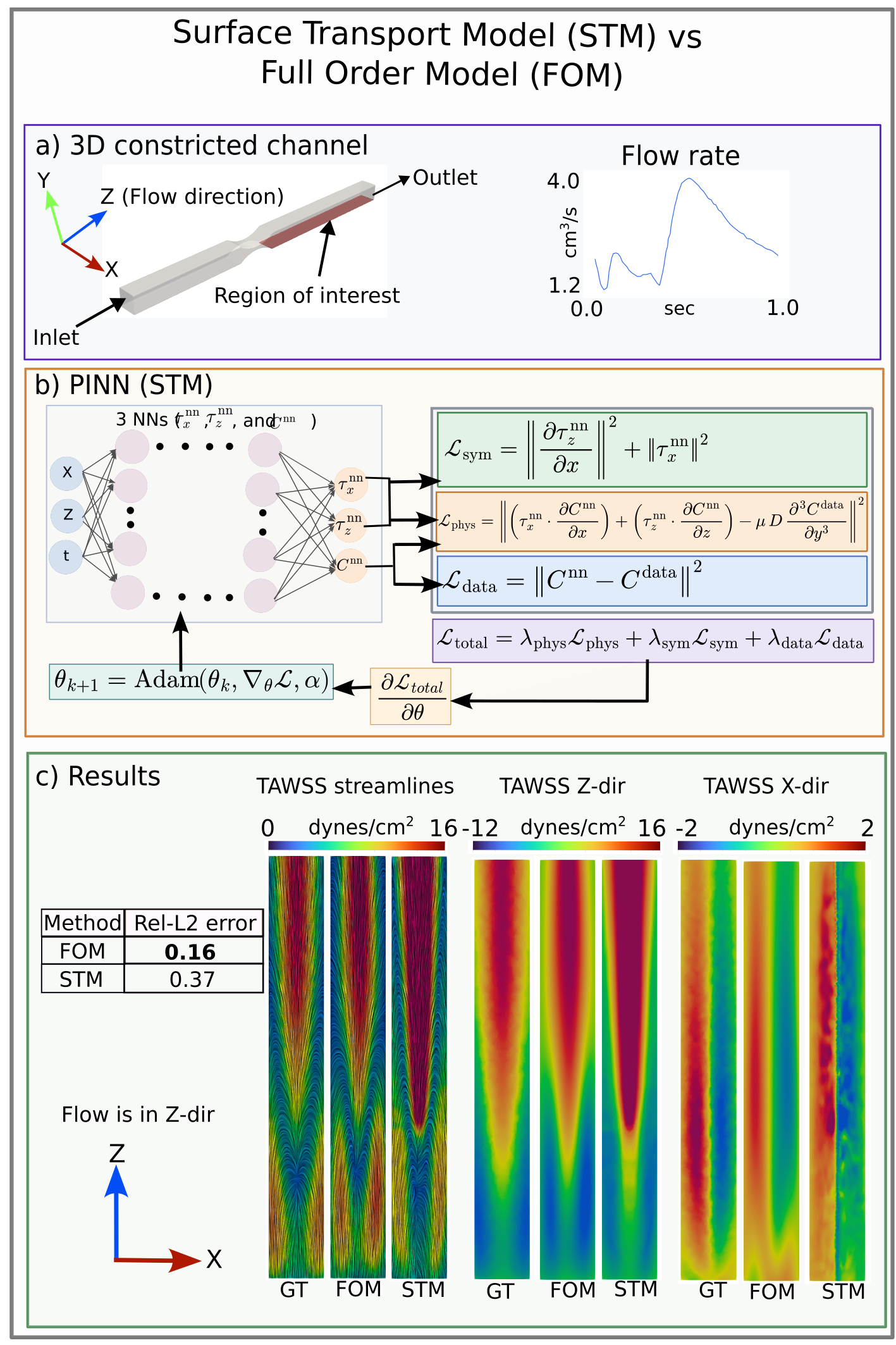}
    \caption{
    Surface Transport Model (STM) setup and results. (a) 3D constricted channel with a \(0.2~\mathrm{cm} \times 0.2~\mathrm{cm}\) cross-section, \(8~\mathrm{cm}\) length, and \(75\%\) stenosis; the red region shows the surface concentration data used for training. (b) STM PINN architecture, where \((t,x,z)\) is used to predict \(C^{NN}\), \(\tau_x\), and \(\tau_z\). (c) TAWSS streamlines colored by magnitude, along with the Z- and X-component comparisons of TAWSS between GT, FOM, and STM, using the same 10 time snapshots over one cardiac cycle.
    } 
    \label{fig:STM}
\end{figure}

To test the STM approach, we used a 3D constricted channel as a test case with pulsatile (unsteady) flow rate, as shown in Fig.~\ref{fig:STM}a. \edit{A channel test case was used here to facilitate the surface transport analysis, which becomes challenging and beyond the scope of our work in curved surfaces.} The channel has a square cross-section of \(0.2~\mathrm{cm} \times 0.2~\mathrm{cm}\) and a total length of \(8~\mathrm{cm}\). A stenosis with \(75\%\) area reduction is introduced along the channel to create a localized constriction. The CFD data for this case were generated by solving the incompressible unsteady Navier--Stokes and continuity equations using SimVascular~\cite{updegrove2017simvascular}. The computational mesh was generated with a tetrahedral element size of \(0.015~\mathrm{cm}\) and four inflation layers near the wall to better resolve the near-wall velocity gradients needed for WSS evaluation. The flow simulation was advanced using a time-step size of \(10^{-3}~\mathrm{s}\) for three cardiac cycles, and the final cycle was retained for the subsequent scalar-transport simulation.

The velocity field from the final cardiac cycle was then imported into FEniCS and used to solve the unsteady advection--diffusion equation for the passive scalar concentration. The scalar diffusivity was set to \(D = 0.005~\mathrm{cm^2/s}\). The advection--diffusion equation was marched with a smaller time-step size of \(10^{-4}~\mathrm{s}\) for \(20000\) time steps. The last cardiac cycle of the resulting concentration field was then used as the training data for the PINN models. The scalar concentration data were generated from a full 3D simulation using a constant inlet concentration boundary condition, \(C=1\). In training, we used 10 temporal snapshots over one pulsatile cycle. The concentration data used in the STM data loss were taken from the surface region highlighted in red in Fig.~\ref{fig:STM}a. This surface data provides the measured scalar field \(C(x,z,t)\), while the third wall-normal derivative term in the STM equation was estimated separately from the near-wall scalar profile.

For comparison with the STM approach, we also implemented a full-order model (FOM) PINN. In the FOM, separate neural networks are used to approximate the velocity components, pressure, and scalar concentration as functions of the space--time coordinates, \((x,y,z,t)\). Five neural networks were used to predict the three velocity components, pressure, and concentration. The FOM enforces the full 3D incompressible Navier--Stokes equations, the continuity equation, and the unsteady advection--diffusion equation through physics-residual losses. The data loss is applied only to scalar concentration measurements extracted from a small 3D masked region near the wall with a thickness of \(0.02~\mathrm{cm}\). Therefore, unlike the STM, which reconstructs WSS directly from wall-surface scalar information, the FOM first reconstructs the full near-wall local flow and scalar fields and then computes WSS from the predicted near-wall velocity gradients as a post-processing step.

The PINN used for the STM does not reconstruct the full 3D velocity field. Instead, the network takes the surface coordinates and time, \((x,z,t)\), as input and predicts three quantities on the wall. Three neural networks were used: one for concentration and one for each of the WSS components on the wall. The supervised data loss is applied to the predicted surface concentration and compares \(C^{NN}(x,z,t)\) with the scalar concentration extracted from the simulation on the highlighted wall region of interest (Fig.~\ref{fig:STM}):
\begin{equation}
    \mathcal{L}_{data}
    =
    \left\|
    C^{NN}(x,z,t) - C(x,z,t)
    \right\|_2^2 .
\end{equation}
The physics loss is constructed from the surface transport equation:
\begin{equation}
    \mathcal{L}_{phy}
    =
    \left\|
    \tau_x \frac{\partial C^{NN}}{\partial x}
    +
    \tau_z \frac{\partial C^{NN}}{\partial z}
    -
    \mu D
    \left(
    \frac{\partial^3 C}{\partial y^3}
    \right)_w
    \right\|_2^2 .
\end{equation}
Therefore, the STM PINN only uses the scalar concentration on the wall surface for the data loss and enforces the reduced surface transport equation as the governing physics constraint. This is different from the full-order PINN (FOM) approach, which requires reconstructing or enforcing the full 3D flow physics in the near-wall region.

The full loss used for the STM training is written as
\begin{equation}
    \mathcal{L}
    =
     \lambda_{phy} \mathcal{L}_{phy}
    +
    \lambda_{data} \mathcal{L}_{data}
    +
    \lambda_{sym} \mathcal{L}_{sym},
\end{equation}
where \(\mathcal{L}_{sym}\) is an additional symmetry loss used to regularize the under-determined WSS reconstruction near the channel centerline. The PINN architecture and the physics loss used are illustrated in Fig.~\ref{fig:STM}b.

This surface transport equation introduces two main challenges. The first challenge is that it provides only one equation for two unknown WSS components, $\tau_x$ and $\tau_z$. Therefore, the problem is under-determined and requires additional constraints or regularization. In this work, we used prior physical knowledge about the symmetry of the flow near the middle of the wall. Specifically, we enforced the lateral WSS component and the lateral variation of the main WSS component to be small at the wall centerline, which was added as a symmetry loss for PINN training:
\begin{equation}
    \tau_x \approx 0 \;,
    \qquad
    \frac{\partial \tau_z}{\partial x} \approx 0 \;. 
\end{equation}

For the FOM, the symmetry loss is also included, but it is applied differently. Since the FOM reconstructs the full velocity field, the symmetry constraint is imposed on the velocity magnitude rather than directly on the WSS components. Specifically, at selected centerline points, the model penalizes the lateral derivative of the velocity magnitude:
\begin{equation}
    \mathcal{L}_{\mathrm{sym}}^{\mathrm{FOM}}
    =
    \left\|
    \frac{\partial |\mathbf{u}|}{\partial x}
    \right\|_2^2 \;.
\end{equation}

The second challenge is the third wall-normal derivative term $(\frac{\partial^3 C}{\partial y^3})_w$ in the STM model, which is difficult to estimate accurately from discrete concentration measurements. We first tested a finite-difference approximation, but the result was noisy and sensitive to numerical error. A more stable approach was to fit the near-wall concentration profile using a third-degree polynomial:
\begin{equation}
    C(y)
    =
    a_0 + a_1 y + a_2 y^2 + a_3 y^3 .
\end{equation}
The third derivative at the wall is then obtained analytically as
\begin{equation}
    \frac{\partial^3 C}{\partial y^3}
    =
    6a_3 \;.
\end{equation}
Therefore, the surface transport relation becomes
\begin{equation}
    \tau_x \frac{\partial C}{\partial x}
    +
    \tau_z \frac{\partial C}{\partial z}
    =
    6 \mu D a_3 .
\end{equation}

This derivative-estimation step also highlights the difference in data requirements between the STM and FOM. For the direct 2D STM equation, the polynomial fitting only requires scalar concentration samples along the wall-normal \(y\)-direction in order to estimate \(\left(\partial^3 C/\partial y^3\right)_w\). In the PINN-based STM implementation, the network only requires scalar concentration data on the wall surface, together with the precomputed or fitted wall-normal derivative term. In contrast, the full-order model requires information from the 3D near-wall region because it reconstructs the flow field and then computes WSS from the near-wall velocity gradients. Therefore, the STM reduces the data requirement from a 3D near-wall flow reconstruction problem to a surface-based scalar transport problem.

Although polynomial fitting reduced the noise compared with finite differences, the reconstructed WSS field still showed noticeable error. This is especially clear in Fig.~\ref{fig:STM}c for the STM results in the \(x\)-direction. We predicted the instantaneous WSS over 10 time snapshots across one cardiac cycle, and the corresponding time-average WSS (TAWSS) results are shown in Fig.~\ref{fig:STM}c. The same number of time snapshots was used for the FOM comparison to keep the evaluation consistent between the two approaches. Although the FOM achieves lower Rel-$L_2$ error, the STM offers two key advantages. First, it bypasses the need to reconstruct the full velocity field, directly relating scalar measurements to WSS. Second, it requires solving only a single surface equation instead of enforcing the full 3D equations, significantly reducing computational cost. This makes the STM useful for a quick qualitative estimation of WSS, especially when only surface scalar and surface gradient measurements are available.

The PINN hyperparameters used for the STM and the FOM are summarized in
Table~\ref{tab:stm_fom_hyperparameters}.

\begin{table}[H]
\centering
\caption{PINN hyperparameters used for the STM and FOM. Here, $\lambda_{\mathrm{data}}$ denotes the concentration data-misfit loss weight, $\lambda_{\mathrm{sym}}$ denotes the symmetry loss weight, $\lambda_{\mathrm{BC}}$ denotes the boundary-condition loss weight, $\lambda_{\mathrm{eqn}}^{\mathrm{STM}}$ denotes the STM governing-equation residual loss weight, $\lambda_{\mathrm{cont}}$ denotes the continuity residual loss weight, $\lambda_{\mathrm{mom}}$ denotes the momentum residual loss weight, and $\lambda_{\mathrm{adv-diff}}$ denotes the scalar advection--diffusion residual loss weight.}
\label{tab:stm_fom_hyperparameters}
\begin{tabular}{lcc}
\hline
\textbf{Hyperparameter} & \textbf{STM} & \textbf{FOM} \\
\hline
Hidden neurons & $200$ & $200$ \\
Hidden layers & $10$ & $5$ \\
Activation function & Swish & Adaptive Swish \\
Optimizer & Adam & Adam \\
Initial learning rate & $10^{-3}$ & $10^{-3}$ \\
Epochs & $3000$ & $3000$ \\
Batch size & $3000$ & $1024 \times 8$ \\
Scheduler decay rate & $0.1$ & $0.1$ \\
Scheduler step size & $1000$ & $1000$ \\
$\lambda_{\mathrm{data}}$ & $100$ & $10^{5}$ \\
$\lambda_{\mathrm{sym}}$ & $1$ & $1$ \\
$\lambda_{\mathrm{BC}}$ & -- & $100$ \\
$\lambda_{\mathrm{eqn}}^{\mathrm{STM}}$ & $1$ & -- \\
$\lambda_{\mathrm{cont}}$ & -- & $1$ \\
$\lambda_{\mathrm{mom}}$ & -- & $1$ \\
$\lambda_{\mathrm{adv-diff}}$ & -- & $100$ \\
\hline
\end{tabular}
\end{table}

\subsection{Uncertainty Quantification}
\label{app:uq}
\edit{To assess the uncertainty and reliability of the reconstructions, we performed five independent runs of
each framework under different sources of variability. For PINN, we varied the random seed, which controls
the neural-network weight initialization, the batch-ordering randomization, and the Fourier-frequency
sampling. For differentiable physics, we used five different initializations of the inlet velocity value.
For the coronary case, each differentiable physics run was carried out for 150 iterations. Across the five
runs, the reconstructed WSS Rel-$L_2$ error was $33.46\% \pm 1.55\%$ for PINN and $2.38\% \pm 0.16\%$
for differentiable physics, indicating that both frameworks are stable to these sources of variability and
that the differentiable physics reconstruction remains substantially more accurate and tightly clustered.}

\subsection{Noise and Sparse Data}
\label{app:noise_sparse}
\edit{
To assess the robustness of the two inference approaches under degraded
measurements, we subjected both the PINN and the differentiable physics method in the coronary case to three tests: additive noise, spatial sparsity, and the two combined.
All errors below are reported as the Rel-$L_2$ error of the recovered WSS against the ground truth reference.

We added independent, identically distributed zero-mean Gaussian noise to the
observed concentration, with standard deviation
$\sigma=(\text{noise level})\times\mathrm{std}(c_{\mathrm{obs}})$ evaluated over
the observation mask, and swept the noise level over $1$, $5$, $10$, $15$, $20$, $30$, $40$, $50$, $60$, and $70\%$. For the
differentiable physics method, noise was applied to the single observed snapshot
(the observed snapshot at t = 0.05 s) and the data misfit was taken as
\begin{equation}
    J(\mathbf{u}_{\mathrm{in}})
    =
    \int_{\Omega_{\mathrm{obs}}}
    \left(
        C(\mathbf{x};\mathbf{u}_{\mathrm{in}})
        - C^{\mathrm{noisy}}(\mathbf{x})
    \right)^{2}
    \, d\Omega ,
\end{equation}
where $C^{\mathrm{noisy}}(\mathbf{x}) = C^{\mathrm{data}}(\mathbf{x}) + \eta(\mathbf{x})$ is
the measured concentration corrupted by zero-mean Gaussian noise
$\eta(\mathbf{x}) \sim \mathcal{N}\!\left(0,\, \sigma^{2}\right)$, with standard deviation
$\sigma$ set as above.

Owing to the computational cost of the adjoint solves, a single run was performed per
noise level. For the PINN, noise was applied to the training concentration data
and three realizations (different seeds) were run per level,
reported as mean\,$\pm$\,std.

As shown in Fig.~\ref{fig:noise_sweep}, the differentiable physics method is
nearly noise-invariant, degrading only from $2.53\%$ to $3.00\%$ across the
entire range. The PINN error remains roughly flat up to $\sim\!20\%$ noise
($\sim\!32$--$36\%$) and then ramps, reaching $57.83\%$ at $40\%$ noise
and $71.26\%$ at $70\%$ noise.

To test robustness to scattered observations, a random fraction of the
observation locations was retained and the
remainder discarded. The data misfit was
evaluated only over the retained data
points. With $10\%$ of the observations kept, the WSS
error rises to $59.77\%$ for the PINN and $5.29\%$ for the differentiable physics
method. Relative to the clean-data baseline ($\sim\!31\%$ and $\sim\!2.5\%$,
respectively), extreme sparsity roughly doubles both errors.

Applying both perturbations together ($10\%$ Gaussian noise with only $10\%$ of
observations retained) yields a WSS error of $63.11\%$ for the PINN and
$5.41\%$ for the differentiable physics method. This combined case is
essentially indistinguishable from sparsity alone (PINN: $63.11\%$ vs.\
$59.77\%$; differentiable physics: $5.41\%$ vs.\ $5.29\%$), confirming that
extreme sparsity dominates compared to noise.

\begin{figure}[h!]
    \centering
    \includegraphics[width=\textwidth]{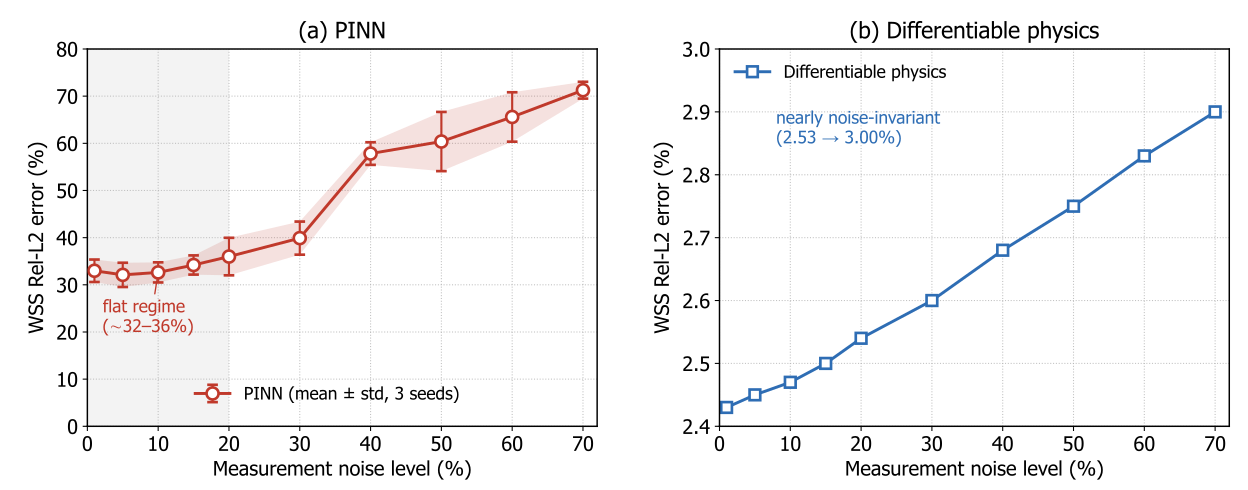}
    \caption{\edit{WSS Rel-$L_2$ error versus measurement noise level.
    (a) PINN (mean\,$\pm$\,std over three seeds) stays flat to $\sim\!20\%$
    noise. (b) The differentiable physics method
    (single run per noise level) degrades only marginally, from $2.53\%$ to
    $3.00\%$. Note the difference in vertical scale between the two panels.}}
    \label{fig:noise_sweep}
\end{figure}

}

\subsection{Sensitivity Analysis}
\label{app:sensitivity}

\edit{
To examine how each method responds to imperfect modeling assumptions and to the
size of the observation region, and to probe the influence of the temporal
sampling used by the differentiable physics method, we applied both the PINN and
the differentiable physics method to the 2D-BFS case (near-wall region) under the following controlled perturbations:
\begin{itemize}
    \item \textbf{Incorrect diffusivity.} The molecular diffusivity was set to twice its true value, $D = 2\times10^{-3}~\mathrm{cm^2/s}$ instead of
the correct $D = 1\times10^{-3}~\mathrm{cm^2/s}$.
    \item \textbf{Reduced region-of-interest thickness.} The observation region
    was halved, from a thickness of $0.1~\mathrm{cm}$ to $0.05~\mathrm{cm}$.
    \item \textbf{Multiple observation snapshots (differentiable physics).}
    Instead of assimilating only the final frame, three snapshots at
    $t=2$, $3$, and $4~\mathrm{s}$ were used. Separately, the sensitivity to the choice of sampling time for the differentiable physics was assessed on the coronary artery case, where an earlier temporal sample at $0.03~\mathrm{s}$ was
used in place of $0.05~\mathrm{s}$.
\end{itemize}

Because the ensemble strategy used elsewhere in this work is
expensive for the differentiable physics solver, we did not repeat these runs
over multiple initializations. Instead, each differentiable physics case was run once for
$150$ iterations, initialized from a uniform inlet guess of
$u_{\mathrm{in}}=1~\mathrm{cm/s}$. Under this restricted budget, the single-run
baseline yielded a WSS Rel-$L_2$ error of $12.89\%$. The PINN retained the
same baseline configuration described previously, achieving a Rel-$L_2$
error of $2.7\%$.

The results reveal a sharp contrast in robustness between the two methods.
Under an \textbf{incorrect diffusivity}, the PINN error grew to to $144.6\%$. The differentiable physics method was affected
far less severely, though its error still nearly doubled, rising from the
$12.89\%$ baseline to $22.3\%$.

Reducing the \textbf{region-of-interest thickness} exposed a similar disparity.
The PINN error increased to $66\%$, indicating a strong sensitivity to the size
of the observation window, whereas the differentiable physics result was almost
unchanged, increasing by only $0.4\%$.

By contrast, assimilating \textbf{multiple observation snapshots} left the
differentiable physics estimate essentially unchanged, with a WSS error of
$12.9\%$, indistinguishable from the single-snapshot baseline.

Finally, for the coronary artery case, moving to the earlier temporal
sample raised the error only slightly, from the $2.5\%$ baseline to $2.96\%$,
indicating that earlier sampling degrades the reconstruction only marginally.

In short, these tests show that the differentiable physics method is
 more robust to modeling and configuration errors than the PINN approach.
}

\subsection{Computational Cost}
\label{app:comp_cost}
\edit{Neither framework is intended for real-time reconstruction in its present form and both operate as offline, with differentiable physics incurring the higher per-iteration cost of a forward and adjoint solve.

Table~\ref{tab:comp_cost} reports the wall-clock cost of the
coronary artery (3D) case. The differentiable physics solver was run on a single core of an AMD EPYC~9654P (Zen~4) CPU, while the
PINN was trained on a single NVIDIA L40S (48~GB) GPU. Although the differentiable physics approach incurs higher costs, as shown in the table, it should be noted that this evaluation utilized only a single core due to limitations in FEniCS adjoint.

\begin{table}[htbp]
\centering
\caption{Computational cost of the coronary artery (3D) reconstruction.}
\label{tab:comp_cost}
\begin{tabular}{lll}
\hline
Quantity & Differentiable physics & PINN \\
\hline
Device               & 1 core, AMD EPYC 9654P (Zen 4) & 1$\times$ NVIDIA L40S (48 GB) \\
Epochs / Iterations  & 300 & 6{,}000 \\
Wall time / epoch-iteration    & 27.5 min & $\approx 7.9$ s \\
One-time setup       & NS ramp $\sim$11--12 min & --- \\
Full-run wall time   & $\approx 137.7$ h ($\approx 5.7$ days) & $\approx 13.2$ h \\
Peak memory          & 7.1 GB RAM & $\sim$6.5 GB GPU \\
\hline
\end{tabular}
\end{table}

%The two methods occupy different cost regimes. The DP solver carries a high per-iteration cost—$27.5$ minutes per epoch for a forward and adjoint solve—and runs for $300$ epochs, for a total of about $137.7$ hours ($5.7$ days) plus an $11$--$12$ minute Navier--Stokes ramp for initialization. The PINN has a much lower per-epoch cost ($\approx 7.9$ s) but requires $6{,}000$ epochs, completing in about $13.2$ hours. Peak memory was comparable in absolute terms ($7.1$ GB host RAM for DP versus $\sim$6.5 GB of GPU memory for the PINN), though on markedly different hardware.

}

\subsection{Experimental Validation}
\label{app:experimental}
\edit{

While the preceding sections demonstrated mostly cases in which the
differentiable physics method is superior, here we provide an example in which
the differentiable physics method cannot be applied because the concentration
boundary condition is missing. To experimentally validate our method for
inferring WSS from scalar measurements, we used the publicly available
experimental dataset provided by Sonnenwald et al.~\cite{Sonnenwald2023}. The
dataset contains dye-visualization measurements of food coloring transported by
steady water flow through a circular pipe. The experiments were conducted over
a range of Reynolds numbers; in the present study, we selected the case
corresponding to $Re=700$.

\paragraph{Flow configuration.}
The pipe diameter is $2.4~\mathrm{cm}$, and the length of the visualized
section was $21.5~\mathrm{cm}$. The dye was injected approximately
$140~\mathrm{cm}$ upstream of the visualized region. When inspecting the
concentration pattern in the dataset, we noticed a small skewness that renders
the concentration field not fully axisymmetric. We attribute this to either a
gravity/inertial effect or a skewed (unknown) inlet concentration boundary condition, which motivates the use of PINN rather than differentiable physics solver.

\paragraph{Analytical reference values.}
The bulk velocity was calculated from the Reynolds number using the density and
dynamic viscosity of water, $\rho=0.998~\mathrm{g/cm^3}$ and
$\mu=0.01~\mathrm{g\,cm^{-1}\,s^{-1}}$, giving
$U_{\mathrm{bulk}}\approx 2.92~\mathrm{cm/s}$. Since the flow is steady and
fully developed, the radial and azimuthal velocity components vanish,
$u_r=u_\theta=0$, and the axial velocity depends on the radial coordinate
alone. The Hagen--Poiseuille profile is
\begin{equation}
u_z(r)=2\,U_{\mathrm{bulk}}\left(1-\frac{r^{2}}{R^{2}}\right),
\end{equation}
giving a centerline (peak) velocity $u_{\max}=2U_{\mathrm{bulk}}$. Using these
parameters, the analytical reference values used for validation are
$U_{\mathrm{bulk}}=2.92~\mathrm{cm/s}$ and
$\tau_w\approx0.0974~\mathrm{dynes/cm^2}$.

\paragraph{PINN inference.}
Because the inlet concentration boundary condition is not reported in the
dataset and due to the observed skewness in the data, a fully differentiable physics approach that marches the transport
equation from a prescribed inlet condition is not feasible here. We therefore
adopted PINN to infer the WSS. The transport of the dye concentration $c$ is
governed by the unsteady advection--diffusion equation, which in cylindrical
coordinates $(r,\theta,z)$ can be written with respect to $U_{\mathrm{bulk}}$.
Substituting $u_r=u_\theta=0$ together with the parabolic axial profile yields
the residual
\begin{equation}
\mathcal{R}\;=\;
\frac{\partial C^{\mathrm{pred}}}{\partial t}
+2\,U_{\mathrm{bulk}}\!\left(1-\frac{r^{2}}{R^{2}}\right)\!
\frac{\partial C^{\mathrm{pred}}}{\partial z}
-\mathcal{D}\left[
\frac{1}{r}\frac{\partial}{\partial r}\!\left(r\frac{\partial C^{\mathrm{pred}}}{\partial r}\right)
+\frac{1}{r^{2}}\frac{\partial^{2} C^{\mathrm{pred}}}{\partial \theta^{2}}
+\frac{\partial^{2} C^{\mathrm{pred}}}{\partial z^{2}}
\right].
\end{equation}
The network $C_\theta(t,z,r)\!\to\!C^{\mathrm{pred}}$ was trained against two
objectives: a data term that fits the measured concentration field of
Fig.~\ref{fig:exp_reconstruction}a, and a physics term that enforces this
residual, with the dye diffusivity fixed at
$\mathcal{D}=5\times10^{-6}~\mathrm{cm^2/s}$. No boundary conditions were
imposed, since the reconstruction domain lies away from the wall. The total
objective is the sum of the concentration data term and the PDE residual term
\begin{equation}
\mathcal{L}=\mathcal{L}_{\mathrm{data}}+\mathcal{L}_{\mathrm{PDE}} \;.
\end{equation}

%where $C^{\mathrm{data}}_i$ are the measured concentrations and $C^{\mathrm{pred}}_i$ the corresponding network predictions.

The training data spanned $t=2.4$--$4.8~\mathrm{s}$ in steps of
$0.08~\mathrm{s}$; this window, step size, and end time gave the best
reconstruction among the settings tested. The concentration network takes the
coordinates $(t,z,r)$ as inputs and passes them through a fixed Fourier-feature
layer with $12$ frequencies. Five fully connected layers of width $96$ with
adaptive Swish activations (per-neuron trainable slope) feed a linear output layer. Only the scalar inverse unknown $U_{\mathrm{bulk}}$
is treated as an additional trainable physical parameter. The weights
were optimized with AdamW (learning rate $2\times10^{-3}$) under a ReduceLROnPlateau schedule ($\times0.5$ on
stall, floor $10^{-6}$), for $2000$ epochs.

\paragraph{Results.}
The PINN concentration predictions are shown in
Fig.~\ref{fig:exp_reconstruction}b. The experimental data are noisy, and the
reconstruction accordingly exhibits a somewhat elevated pointwise error. The final
recovered velocity profile and the developing trajectory of $U_{\mathrm{bulk}}$ during training and are shown in Figs.~\ref{fig:exp_reconstruction}c
and~d. Using the predicted velocity, the inferred WSS is
$\tau_w=0.094~\mathrm{dynes/cm^2}$, a relative error of $3.5\%$
(underprediction) against the analytical reference. Although the measured
concentration is not fully axisymmetric---for the reasons noted above (a
skewed, unknown boundary condition or a gravity effect)---the PINN still
performs well as it solves the problem locally and over a short time
window. 
%This demonstrates the robustness and superiority of PINN in real-world test cases, reconstructing the WSS even from noisy concentration data.

\begin{figure}[h!]
    \centering
    \includegraphics[width=0.7\textwidth, height=\textheight, keepaspectratio]{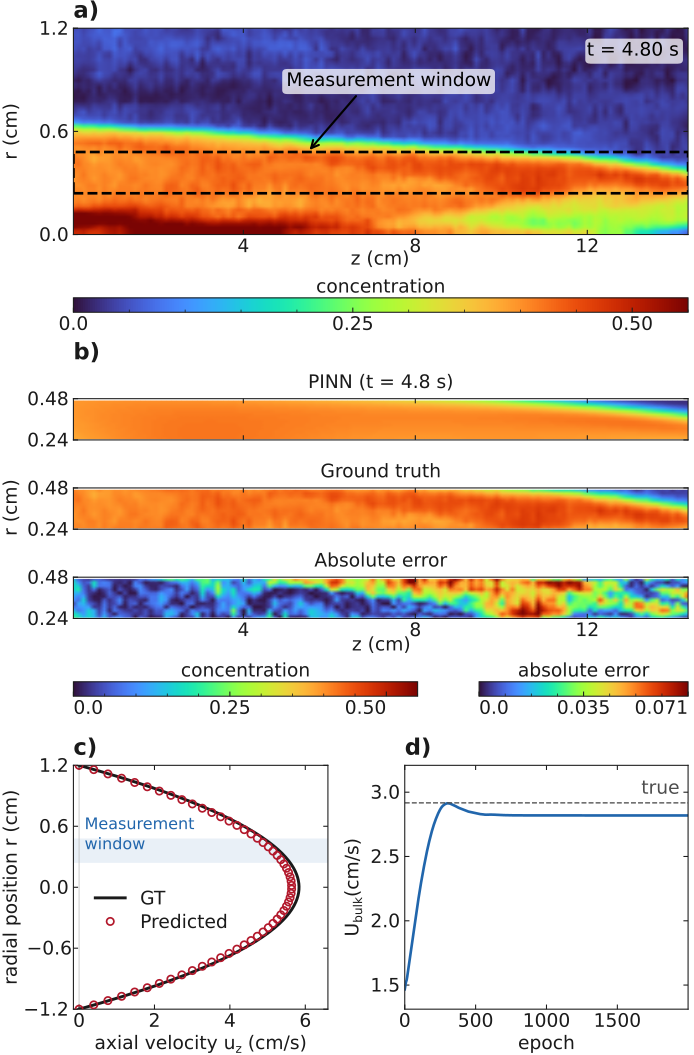}
    \caption{\edit{In vitro reconstruction of the velocity measurements from a
    dye tracer inside a pipe using PINN. The dye concentration is observed
    only within a narrow off-axis measurement window, and the network must
    reconstruct the surrounding field and back out the velocity that transports
    it.
    (a)~Reference dye concentration over the $r$--$z$ plane; the dashed box
    outlines the measurement window supplied to the network.
    (b)~Within the measurement window, side-by-side view of the network's
    prediction, the reference field, and the discrepancy between them.
    (c)~Radial profile of the axial velocity returned by the network (markers)
    overlaid on the reference profile (solid line), with the measurement window
    shaded.
    (d)~Training history of the estimated bulk velocity as it settles onto the
    reference value (dashed).}}
    \label{fig:exp_reconstruction}
\end{figure}

}

\subsection{Mesh Independence Study}
\label{app:mesh}

\edit{Refining the 2D-BFS mesh from $37{,}838$ to $151{,}352$ cells changed the mean WSS from $0.1118$ to $0.1128~\mathrm{dynes/cm^2}$, corresponding to a relative difference of $0.9\%$. Similarly, refining the 3D coronary artery mesh from $688{,}865$ to $1{,}228{,}432$ cells changed the mean WSS from $56.46$ to $58.09~\mathrm{dynes/cm^2}$, corresponding to a relative difference of $2.81\%$. These small differences indicate that the meshes used in this work yield sufficiently mesh-independent results.}

\subsection{Practitioner Guidance: PINN vs.\ Differentiable Physics}
\label{app:guidance}

\edit{Table~\ref{tab:pinn_dp_guidance} summarizes the practical trade-offs between
the PINN (soft-constraint) and differentiable physics / discrete adjoint
(hard-constraint) frameworks, to guide method selection for WSS reconstruction
from passive scalar observations.}

\edit{
\begin{table}[h!]
\centering
\caption{Practitioner guidance: comparison of PINN (soft-constraint) and
differentiable physics / discrete adjoint (hard-constraint) frameworks for
WSS reconstruction from passive scalar observations.}
\label{tab:pinn_dp_guidance}
\renewcommand{\arraystretch}{1.25}
\resizebox{\textwidth}{!}{%
\begin{tabular}{lll}
\hline
\textbf{Attribute}
& \textbf{PINN (soft constraints)}
& \textbf{Differentiable physics / discrete adjoint (hard constraints)} \\
\hline
Governing equations
& Penalized as residual loss (approximate)
& Enforced exactly (to discretization accuracy) \\

Optimization variable
& Network weights for $\mathbf{u}$, $p$, and $C$
& Inlet velocity profile (boundary control) \\

Iterative forward solve
& Not required
& Required at every iteration \\

Full domain/geometry
& Not required (local reconstruction possible)
& Required \\

Scalar boundary conditions
& Not required
& Required \\

Discretization
& Mesh-free (collocation points)
& FEM mesh \\

Cost per iteration
& Low (network forward/backward pass)
& High (forward $+$ adjoint solve) \\

Sensitivity to data location
& High
& Lower \\

Sensitivity to hyperparameters
& High (loss weights and sampling)
& Low \\

Sensitivity to initialization
& Lower
& High (motivates an ensemble strategy) \\

WSS accuracy in this study
&Accurate for near-wall 2D observations; poor quantitative accuracy for far-field 2D and the 3D coronary case
& Robust across all observation scenarios \\
\hline

\multicolumn{3}{l}{\textbf{Preferred when}} \\

Geometry or boundary conditions are incomplete
& \checkmark
& \\

Local near-wall reconstruction is sufficient
& \checkmark
& \\

A forward numerical model is unavailable
& \checkmark
& \\

Low per-iteration computational cost is important
& \checkmark
& \\

A validated forward model is available
&
& \checkmark \\

Scalar boundary conditions are known
&
& \checkmark \\

Global reconstruction is required
&
& \checkmark \\

High physical fidelity is paramount
&
& \checkmark \\
\hline
\end{tabular}%
}
\end{table}
}

\subsection{Correlation Between Observable Error and WSS Accuracy}
\label{app:conc_wss}

\begin{figure}[h!]
    \centering
    \includegraphics[width=.85\textwidth]{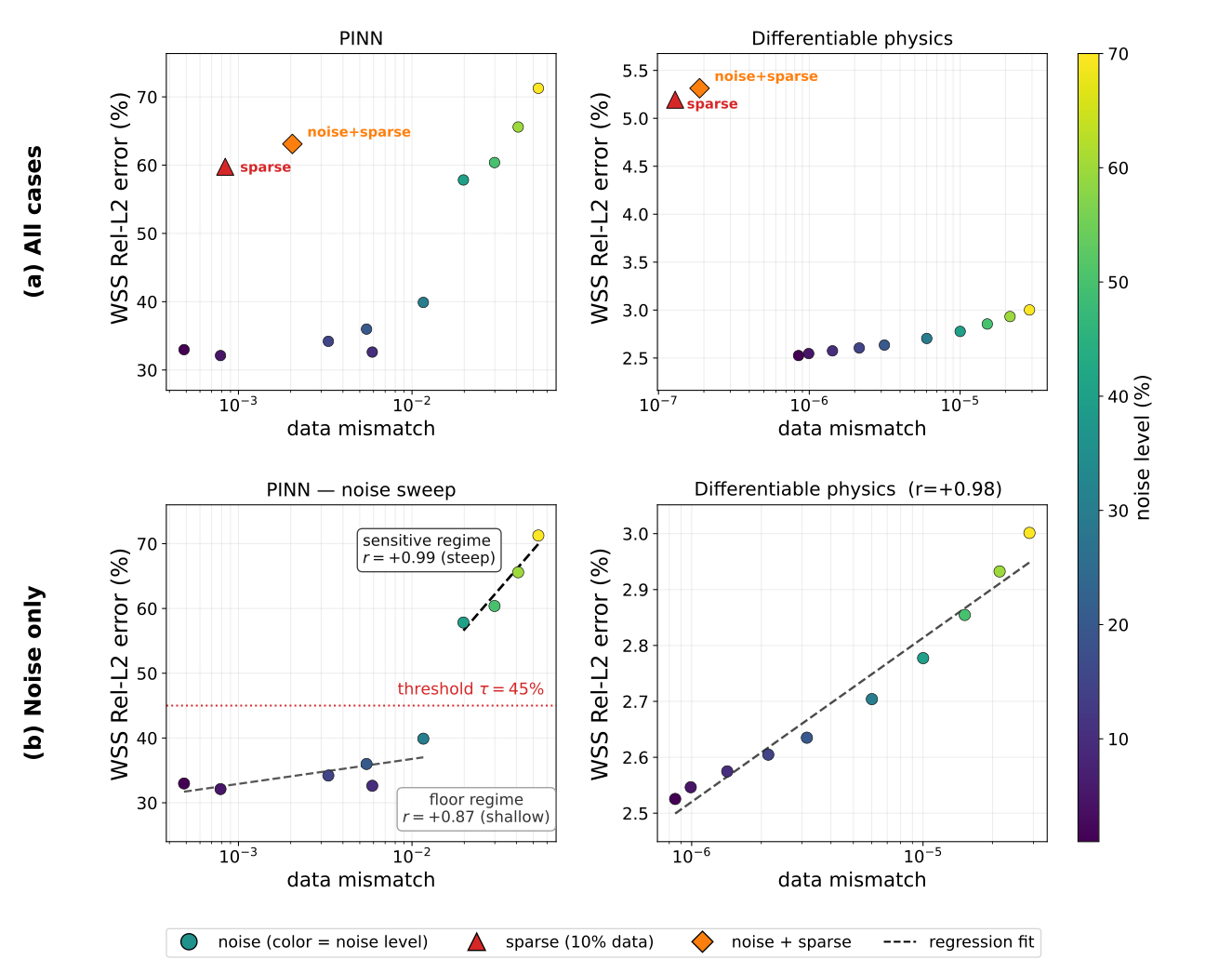}
    \caption{\edit{Data mismatch (concentration data loss, horizontal axis) versus WSS
    Rel-$L_2$ error (vertical axis) for the coronary example. PINN (left column) and the differentiable physics
    method (right column) are shown. Marker shape shows the type of perturbation: circles are noise
    cases (colored by noise level), triangles are sparse cases, and diamonds are combined
    noise-and-sparse cases. PINN points are the mean of three noise runs per level.
    Row (a) shows all cases together; row (b) shows the noise cases only.}}
    \label{fig:conc_wss_corr}
\end{figure}

\edit{In practical applications, the true WSS is unavailable; therefore, reconstruction accuracy must be assessed using observable quantities. One such quantity is the data mismatch, defined as the discrepancy between the reconstructed and measured concentration fields. Because the concentration field is the measured input, this mismatch can be evaluated without knowledge of the true WSS. To examine whether it can serve as an indicator of WSS accuracy, we compare the data mismatch with the Rel-$L_2$ error in WSS for the coronary artery cases reconstructed using PINN and differentiable physics. The noise, sparsity, and combined noise-and-sparsity cases introduced previously are included in this analysis. We consider two comparisons: all perturbation cases together (Fig.~\ref{fig:conc_wss_corr}a) and the noise-only cases (Fig.~\ref{fig:conc_wss_corr}b). The sparse and combined noise-and-sparsity cases are identified using distinct markers in the all-cases panel. Because the PINN and differentiable physics results have substantially different error ranges, they are shown separately on the left and right, respectively. For PINN, each noise-case value represents the mean of three independent runs.}

\edit{For the differentiable physics method, the data mismatch provides a reliable indicator of WSS accuracy when the observations are corrupted by noise. Across the noise levels considered, the data-mismatch and WSS errors increase together and exhibit a very strong correlation ($r=0.98$; Fig.~\ref{fig:conc_wss_corr}b). Thus, within the noise-only cases, a smaller data mismatch generally corresponds to a more accurate WSS reconstruction. For PINN, the relationship is also strong but exhibits two distinct regimes around a WSS error of approximately $45\%$. Above this value, the data mismatch increases steeply with the WSS error with a strong correlation ($r=0.99$). Below this value, the correlation remains strong ($r=0.87$), but the data mismatch changes less consistently and is therefore less sensitive to variations in WSS accuracy. Consequently, the data mismatch remains informative for PINN, although it provides a clearer indication of reconstruction quality in the higher-error regime.}

\edit{The results further suggest that the data mismatch may remain informative when the sparsity level is held fixed. In particular, the case with $10\%$ sparsity and $10\%$ noise is consistent with the relationship between data-mismatch and WSS errors observed for the noise-only cases when compared with the corresponding noise-free case at $10\%$ sparsity. This observation suggests that increasing noise may produce a corresponding increase in both errors when the amount of available data is unchanged. However, because only one noisy case was evaluated at the $10\%$ sparsity level, this interpretation remains a hypothesis. Additional experiments involving multiple noise levels and independent noise realizations at fixed sparsity are required to confirm this relationship.}

\clearpage

\bibliography{refs}

\begin{thebibliography}{10}

\bibitem{tarbell2010shear}
J.~M. Tarbell.
\newblock Shear stress and the endothelial transport barrier.
\newblock {\em Cardiovascular Research}, 87(2):320--330, 2010.

\bibitem{conway2013flow}
D.~E. Conway and M.~A. Schwartz.
\newblock Flow-dependent cellular mechanotransduction in atherosclerosis.
\newblock {\em Journal of Cell Science}, 126(22):5101--5109, 2013.

\bibitem{chatterjee2018endothelial}
S.~Chatterjee.
\newblock Endothelial mechanotransduction, redox signaling and the regulation of vascular inflammatory pathways.
\newblock {\em Frontiers in Physiology}, 9:524, 2018.

\bibitem{meng2014high}
H.~Meng, V.~M. Tutino, J.~Xiang, and A.~Siddiqui.
\newblock High {WSS} or low {WSS}? complex interactions of hemodynamics with intracranial aneurysm initiation, growth, and rupture: toward a unifying hypothesis.
\newblock {\em American Journal of Neuroradiology}, 35(7):1254--1262, 2014.

\bibitem{morel2021effects}
S.~Morel, S.~Schilling, M.~R. Diagbouga, M.~Delucchi, M.-L. Bochaton-Piallat, S.~Lemeille, S.~Hirsch, and B.~R. Kwak.
\newblock Effects of low and high aneurysmal wall shear stress on endothelial cell behavior: differences and similarities.
\newblock {\em Frontiers in Physiology}, 12:727338, 2021.

\bibitem{de2024predicting}
G.~De~Nisco, E.~M.~J. Hartman, E.~Torta, J.~Daemen, C.~Chiastra, D.~Gallo, U.~Morbiducci, and J.~J. Wentzel.
\newblock Predicting lipid-rich plaque progression in coronary arteries using multimodal imaging and wall shear stress signatures.
\newblock {\em Arteriosclerosis, Thrombosis, and Vascular Biology}, 44(4):976--986, 2024.

\bibitem{mahmoudi2021story}
M.~Mahmoudi, A.~Farghadan, D.~R. McConnell, A.~J. Barker, J.~J. Wentzel, M.~J. Budoff, and A.~Arzani.
\newblock The story of wall shear stress in coronary artery atherosclerosis: biochemical transport and mechanotransduction.
\newblock {\em Journal of Biomechanical Engineering}, 143(4):041002, 2021.

\bibitem{schlichting1961boundary}
H.~Schlichting and J.~Kestin.
\newblock {\em Boundary layer theory}, volume 121.
\newblock Springer, 1961.

\bibitem{katritsis2007wall}
D.~Katritsis, L.~Kaiktsis, A.~Chaniotis, J.~Pantos, E.~P. Efstathopoulos, and V.~Marmarelis.
\newblock Wall shear stress: theoretical considerations and methods of measurement.
\newblock {\em Progress in cardiovascular diseases}, 49(5):307--329, 2007.

\bibitem{wang2020predicting}
H.~Wang, Z.~Yang, B.~Li, and S.~Wang.
\newblock Predicting the near-wall velocity of wall turbulence using a neural network for particle image velocimetry.
\newblock {\em Physics of Fluids}, 32(11), 2020.

\bibitem{papaioannou2005vascular}
T.~G. Papaioannou and Christodoulos et~al. Stefanadis.
\newblock Vascular wall shear stress: basic principles and methods.
\newblock {\em Hellenic J. C.}, 46(1):9--15, 2005.

\bibitem{crimaldi2008planar}
J.~P. Crimaldi.
\newblock Planar laser induced fluorescence in aqueous flows.
\newblock {\em Experiments in Fluids}, 44(6):851--863, 2008.

\bibitem{wang2000high}
G.~Wang and H.~Fiedler.
\newblock On high spatial resolution scalar measurement with {LIF} part 1: photobleaching and thermal blooming: part 1: photobleaching and thermal blooming.
\newblock {\em Experiments in fluids}, 29(3):257--264, 2000.

\bibitem{woodworth2023heat}
A.~D. Woodworth, D.~M. Salazar, and T.~Liu.
\newblock Heat transfer and skin friction: beyond the reynolds analogy.
\newblock {\em International Journal of Heat and Mass Transfer}, 206:123960, 2023.

\bibitem{liao2023physics}
X.~Liao, Z.~Cai, J.~Chen, T.~Liu, and J.-H. Lai.
\newblock Physics-based optical flow estimation under varying illumination conditions.
\newblock {\em Signal processing: Image Communication}, 117:117007, 2023.

\bibitem{sharma2019analytic}
A.~Sharma, I.~I. Rypina, R.~Musgrave, and G.~Haller.
\newblock Analytic reconstruction of a two-dimensional velocity field from an observed diffusive scalar.
\newblock {\em Journal of Fluid Mechanics}, 871:755--774, 2019.

\bibitem{su1996scalar}
L.~K. Su and W.~J. Dahm.
\newblock Scalar imaging velocimetry measurements of the velocity gradient tensor field in turbulent flows. i. assessment of errors.
\newblock {\em Physics of Fluids}, 8(7):1869--1882, 1996.

\bibitem{lardo2015estimating}
A.~C. Lardo, A.~A. Rahsepar, J.~H. Seo, P.~Eslami, F.~Korley, S.~Kishi, T.~Abd, R.~Mittal, and R.~T. George.
\newblock Estimating coronary blood flow using {CT} transluminal attenuation flow encoding: Formulation, preclinical validation, and clinical feasibility.
\newblock {\em Journal of Cardiovascular Computed Tomography}, 9(6):559--566, 2015.

\bibitem{eslami2015computational}
P.~Eslami, J.-H. Seo, A.~A. Rahsepar, R.~George, A.~C. Lardo, and R.~Mittal.
\newblock Computational study of computed tomography contrast gradients in models of stenosed coronary arteries.
\newblock {\em Journal of Biomechanical Engineering}, 137(9):091002, 2015.

\bibitem{raissi2020hidden}
M.~Raissi, A.~Yazdani, and G.~E. Karniadakis.
\newblock Hidden fluid mechanics: Learning velocity and pressure fields from flow visualizations.
\newblock {\em Science}, 367(6481):1026--1030, 2020.

\bibitem{gunzburger2002perspectives}
M.~D. Gunzburger.
\newblock {\em Perspectives in flow control and optimization}.
\newblock SIAM, 2002.

\bibitem{giles2000introduction}
M.~B. Giles and N.~A. Pierce.
\newblock An introduction to the adjoint approach to design.
\newblock {\em Flow, Turbulence and Combustion}, 65(3):393--415, 2000.

\bibitem{kenway2019effective}
G.~K. Kenway, C.~A. Mader, P.~He, and J.~R. Martins.
\newblock Effective adjoint approaches for computational fluid dynamics.
\newblock {\em Progress in Aerospace Sciences}, 110:100542, 2019.

\bibitem{mcnamara2004fluid}
A.~McNamara, A.~Treuille, Z.~Popovi{\'c}, and J.~Stam.
\newblock Fluid control using the adjoint method.
\newblock {\em ACM Transactions On Graphics (TOG)}, 23(3):449--456, 2004.

\bibitem{raissi2019physics}
M.~Raissi, P.~Perdikaris, and G.~E. Karniadakis.
\newblock Physics-informed neural networks: A deep learning framework for solving forward and inverse problems involving nonlinear partial differential equations.
\newblock {\em Journal of Computational Physics}, 378:686--707, 2019.

\bibitem{arzani2021uncovering}
A.~Arzani, J.~Wang, and R.~M. D'Souza.
\newblock Uncovering near-wall blood flow from sparse data with physics-informed neural networks.
\newblock {\em Physics of Fluids}, 33(7), 2021.

\bibitem{jagtap2022physics}
A.~D. Jagtap, Z.~Mao, N.~Adams, and G.~E. Karniadakis.
\newblock Physics-informed neural networks for inverse problems in supersonic flows.
\newblock {\em Journal of Computational Physics}, 466:111402, 2022.

\bibitem{kim2024review}
D.~Kim and J.~Lee.
\newblock A review of physics informed neural networks for multiscale analysis and inverse problems.
\newblock {\em Multiscale Science and Engineering}, 6(1):1--11, 2024.

\bibitem{mai2024two}
J.~Mai, Y.~Li, L.~Long, Y.~Huang, H.~Zhang, and Y.~You.
\newblock Two-dimensional temperature field inversion of turbine blade based on physics-informed neural networks.
\newblock {\em Physics of Fluids}, 36(3), 2024.

\bibitem{gaidzik2021hemodynamic}
F.~Gaidzik, S.~Pathiraja, S.~Saalfeld, D.~Stucht, O.~Speck, D.~Th{\'e}venin, and G.~Janiga.
\newblock Hemodynamic data assimilation in a subject-specific circle of willis geometry.
\newblock {\em Clinical Neuroradiology}, 31(3):643--651, 2021.

\bibitem{canuto2020ensemble}
D.~Canuto, J.~L. Pantoja, J.~Han, E.~P. Dutson, and J.~D. Eldredge.
\newblock An ensemble kalman filter approach to parameter estimation for patient-specific cardiovascular flow modeling.
\newblock {\em Theoretical and Computational Fluid Dynamics}, 34(4):521--544, 2020.

\bibitem{kalnay2003atmospheric}
E.~Kalnay.
\newblock {\em Atmospheric modeling, data assimilation and predictability}.
\newblock Cambridge University Press, 2003.

\bibitem{talagrand1987variational}
O.~Talagrand and P.~Courtier.
\newblock Variational assimilation of meteorological observations with the adjoint vorticity equation. i: Theory.
\newblock {\em Quarterly Journal of the Royal Meteorological Society}, 113(478):1311--1328, 1987.

\bibitem{evensen2009data}
G.~Evensen.
\newblock {\em Data assimilation: the ensemble Kalman filter}.
\newblock Springer, 2009.

\bibitem{habibi2021integrating}
M.~Habibi, R.~M. D'Souza, S.~T.~M. Dawson, and A.~Arzani.
\newblock Integrating multi-fidelity blood flow data with reduced-order data assimilation.
\newblock {\em Computers in Biology and Medicine}, 135:104566, 2021.

\bibitem{elhadidy2026sle}
M.~Elhadidy, R.~D'Souza, and A.~Arzani.
\newblock {SLE-FNO}: Single-layer extensions for task-agnostic continual learning in fourier neural operators.
\newblock {\em Physica Scripta}, 2026.

\bibitem{buffoni2001adjoint}
G.~Buffoni and E.~Cupini.
\newblock The adjoint advection-diffusion equation in stationary and time dependent problems: a reciprocity relation.
\newblock {\em Rivista di Matematica Della Universita di Parma}, 4:9--19, 2001.

\bibitem{gillissen2018space}
J.~J. Gillissen, A.~Vilquin, H.~Kellay, R.~Bouffanais, and D.~K. Yue.
\newblock A space--time integral minimisation method for the reconstruction of velocity fields from measured scalar fields.
\newblock {\em Journal of Fluid Mechanics}, 854:348--366, 2018.

\bibitem{bakker2021image}
L.~Bakker, N.~Xiao, A.~Van De~Ven, M.~Schaap, F.~Van De~Vosse, and C.~Taylor.
\newblock Image-based blood flow estimation using a semi-analytical solution to the advection--diffusion equation in cylindrical domains.
\newblock {\em Journal of Fluid Mechanics}, 924:A18, 2021.

\bibitem{liu2021perfusion}
P.~Liu, Y.~Z. Lee, S.~R. Aylward, and M.~Niethammer.
\newblock Perfusion imaging: an advection diffusion approach.
\newblock {\em IEEE Transactions on Medical Imaging}, 40(12):3424--3435, 2021.

\bibitem{huang2021reconstruction}
S.~Huang, M.~Sigovan, and B.~Sixou.
\newblock Reconstruction of vascular blood flow in a vessel from tomographic projections.
\newblock {\em Biomedical Physics \& Engineering Express}, 7(6):065032, 2021.

\bibitem{riemer2022contrast}
K.~Riemer, E.~M. Rowland, J.~Broughton-Venner, C.~H. Leow, M.~Tang, and P.~Weinberg.
\newblock Contrast agent-free assessment of blood flow and wall shear stress in the rabbit aorta using ultrasound image velocimetry.
\newblock {\em Ultrasound in Medicine \& Biology}, 48(3):437--449, 2022.

\bibitem{allen2018extraction}
D.~R. Allen, K.~W. Hoppel, and D.~D. Kuhl.
\newblock Extraction of wind and temperature information from hybrid 4d-var assimilation of stratospheric ozone using navgem.
\newblock {\em Atmospheric Chemistry and Physics}, 18(4):2999--3026, 2018.

\bibitem{shusong2024deep}
H.~Shusong, S.~Monica, and S.~Bruno.
\newblock Deep learning methods for blood flow reconstruction in a vessel with contrast enhanced x-ray computed tomography.
\newblock {\em International Journal for Numerical Methods in Biomedical Engineering}, 40(1):e3785, 2024.

\bibitem{sarabian2022physics}
M.~Sarabian, H.~Babaee, and K.~Laksari.
\newblock Physics-informed neural networks for brain hemodynamic predictions using medical imaging.
\newblock {\em IEEE Transactions on Medical Imaging}, 41(9):2285--2303, 2022.

\bibitem{sierpe2025estimation}
M.~Sierpe, E.~Castillo, H.~Mella, and F.~Galarce.
\newblock Estimation of hemodynamic parameters via physics informed neural networks including hematocrit dependent rheology.
\newblock {\em arXiv preprint arXiv:2508.03326}, 2025.

\bibitem{toscano2025mr}
J.~D. Toscano, Y.~Guo, Z.~Wang, M.~Vaezi, Y.~Mori, G.~E. Karniadakis, K.~A. Boster, and D.~H. Kelley.
\newblock {{MR-AIV} reveals in vivo brain-wide fluid flow with physics-informed {AI}}.
\newblock {\em Science Advances}, 12(22):eaeb0404, 2026.

\bibitem{kalajahi2025input}
A.~P. Kalajahi, H.~Csala, Z.~B. Mamun, S.~Yadav, O.~Amili, A.~Arzani, and R.~M. D’Souza.
\newblock Input parameterized physics informed neural networks for de noising, super-resolution, and imaging artifact mitigation in time resolved three dimensional phase-contrast magnetic resonance imaging.
\newblock {\em Engineering Applications of Artificial Intelligence}, 150:110600, 2025.

\bibitem{thakur2026punch}
S.~Thakur, M.~Roper, Y.~Zhou, R.~A. Bafghi, B.~K. Nallamothu, C.~A. Figueroa, S.~Paruchuri, S.~Burger, and M.~Raissi.
\newblock {PUNCH}: Physics-informed uncertainty-aware network for coronary hemodynamics.
\newblock {\em arXiv preprint arXiv:2601.17192}, 2026.

\bibitem{arzani2018wall}
A.~Arzani and S.~C. Shadden.
\newblock Wall shear stress fixed points in cardiovascular fluid mechanics.
\newblock {\em Journal of Biomechanics}, 73:145--152, 2018.

\bibitem{du2023state}
Y.~Du, M.~Wang, and T.~A. Zaki.
\newblock State estimation in minimal turbulent channel flow: A comparative study of {4DVar} and {PINN}.
\newblock {\em International Journal of Heat and Fluid Flow}, 99:109073, 2023.

\bibitem{mitusch2019dolfin}
S.~Mitusch, S.~Funke, and J.~Dokken.
\newblock dolfin-adjoint {2018.1}: automated adjoints for {FEniCS} and {Firedrake}.
\newblock {\em Journal of Open Source Software}, 4(38):1292, 2019.

\bibitem{arzani2018accounting}
A.~Arzani.
\newblock Accounting for residence-time in blood rheology models: do we really need non-newtonian blood flow modelling in large arteries?
\newblock {\em Journal of The Royal Society Interface}, 15(146), 2018.

\bibitem{fiadeiro1984obtaining}
M.~E. Fiadeiro and G.~Veronis.
\newblock Obtaining velocities from tracer distributions.
\newblock {\em Journal of Physical Oceanography}, 14(11):1734--1746, 1984.

\bibitem{wunsch1985can}
C.~Wunsch.
\newblock Can a tracer field be inverted for velocity?
\newblock {\em Journal of Physical Oceanography}, 15(11):1521--1531, 1985.

\bibitem{loshchilov2017decoupled}
I.~Loshchilov and F.~Hutter.
\newblock Decoupled weight decay regularization.
\newblock {\em arXiv preprint arXiv:1711.05101}, 2017.

\bibitem{kingma2014adam}
D.~P. Kingma and J.~Ba.
\newblock Adam: A method for stochastic optimization.
\newblock {\em arXiv preprint arXiv:1412.6980}, 2014.

\bibitem{ramsundar2021differentiable}
B.~Ramsundar, D.~Krishnamurthy, and V.~Viswanathan.
\newblock Differentiable physics: A position piece.
\newblock {\em arXiv preprint arXiv:2109.07573}, 2021.

\bibitem{blondel2024elements}
M.~Blondel and V.~Roulet.
\newblock The elements of differentiable programming.
\newblock {\em arXiv preprint arXiv:2403.14606}, 2024.

\bibitem{farrell2013automated}
P.~E. Farrell, D.~A. Ham, S.~W. Funke, and M.~E. Rognes.
\newblock Automated derivation of the adjoint of high-level transient finite element programs.
\newblock {\em SIAM Journal on Scientific Computing}, 35(4):C369--C393, 2013.

\bibitem{tancik2020fourier}
M.~Tancik, P.~Srinivasan, B.~Mildenhall, S.~Fridovich-Keil, N.~Raghavan, U.~Singhal, R.~Ramamoorthi, J.~Barron, and R.~Ng.
\newblock Fourier features let networks learn high frequency functions in low dimensional domains.
\newblock {\em Advances in Neural Information Processing Systems}, 33:7537--7547, 2020.

\bibitem{sallam2023use}
O.~Sallam and M.~F{\"u}rth.
\newblock On the use of fourier features-physics informed neural networks (ff-pinn) for forward and inverse fluid mechanics problems.
\newblock {\em Proceedings of the Institution of Mechanical Engineers, Part M: Journal of Engineering for the Maritime Environment}, 237(4):846--866, 2023.

\bibitem{rahimi2007random}
A.~Rahimi and B.~Recht.
\newblock Random features for large-scale kernel machines.
\newblock {\em Advances in Neural Information Processing Systems}, 20, 2007.

\bibitem{farghadan2019combined}
A.~Farghadan and A.~Arzani.
\newblock The combined effect of wall shear stress topology and magnitude on cardiovascular mass transport.
\newblock {\em International Journal of Heat and Mass Transfer}, 131:252--260, 2019.

\bibitem{mirramezani2019reduced}
M.~Mirramezani, S.~L. Diamond, H.~I. Litt, and S.~C. Shadden.
\newblock Reduced order models for transstenotic pressure drop in the coronary arteries.
\newblock {\em Journal of Biomechanical Engineering}, 141(3):031005, 2019.

\bibitem{updegrove2017simvascular}
A.~Updegrove, N.~M. Wilson, J.~Merkow, H.~Lan, A.~L. Marsden, and S.~C. Shadden.
\newblock {SimVascular}: an open source pipeline for cardiovascular simulation.
\newblock {\em Annals of Biomedical Engineering}, 45(3):525--541, 2017.

\bibitem{arzani2016lagrangian}
A.~Arzani, A.~M. Gambaruto, G.~Chen, and S.~C. Shadden.
\newblock Lagrangian wall shear stress structures and near-wall transport in high-schmidt-number aneurysmal flows.
\newblock {\em Journal of Fluid Mechanics}, 790:158--172, 2016.

\bibitem{hansen2016reduced}
K.~B. Hansen and S.~C. Shadden.
\newblock A reduced-dimensional model for near-wall transport in cardiovascular flows.
\newblock {\em Biomechanics and Modeling in Mechanobiology}, 15(3):713--722, 2016.

\bibitem{chen2019relations}
T.~Chen, T.~Liu, L.~P. Wang, and S.~Chen.
\newblock Relations between skin friction and other surface quantities in viscous flows.
\newblock {\em Physics of Fluids}, 31(10), 2019.

\bibitem{ohashi2026physics}
N.~Ohashi, H.~Cao, L.~K. Hwang, P.~K. Kang, and B.~Kwon.
\newblock Physics-informed neural networks with domain decomposition for inferring velocity fields from concentration fields.
\newblock {\em AI Thermal Fluids}, page 100037, 2026.

\bibitem{rawden2026physics}
J.~I. Rawden, C.~Vanderwel, and S.~Symon.
\newblock Physics-informed neural networks for passive scalar emission and transport.
\newblock {\em Physical Review Fluids}, 11(2):024501, 2026.

\bibitem{thawon2026physics}
I.~Thawon, D.~Vo, T.~Q. Bui, K.~Rattanamongkhonkun, C.~Chamroon, N.~Tippayawong, Y.~Mona, R.~Wanison, and P.~Suttakul.
\newblock Physics-informed neural networks: Current progress and challenges in computational solid and structural mechanics.
\newblock {\em Computer Modeling in Engineering \& Sciences {(CMES)}}, 146(2):1, 2026.

\bibitem{jnini2026curvature}
A.~Jnini, E.~Kiyani, K.~Shukla, J.~F. Urban, N.~Ahmadi Daryakenari, J.~Muller, M.~Zeinhofer, and G.~E. Karniadakis.
\newblock Curvature-aware optimization for high-accuracy physics-informed neural networks.
\newblock {\em arXiv preprint arXiv:2604.05230}, 2026.

\bibitem{wang2025simulating}
S.~Wang, S.~Sankaran, X.~Fan, P.~Stinis, and P.~Perdikaris.
\newblock Simulating three-dimensional turbulence with physics-informed neural networks.
\newblock {\em arXiv preprint arXiv:2507.08972}, 2025.

\bibitem{costabal2024delta}
F.~S. Costabal, S.~Pezzuto, and P.~Perdikaris.
\newblock {$\Delta$-PINNs}: Physics-informed neural networks on complex geometries.
\newblock {\em Engineering Applications of Artificial Intelligence}, 127:107324, 2024.

\bibitem{yang2026assessing}
H.~Yang, J.~Zhang, B.~K. Nallamothu, K.~Garikipati, and C.~A. Figueroa.
\newblock Assessing coronary microvascular dysfunction using angiography-based data-driven methods.
\newblock {\em Computer Methods in Applied Mechanics and Engineering}, 452:118743, 2026.

\bibitem{Sonnenwald2023}
F.~Sonnenwald, Z.~Peng, and I.~Guymer.
\newblock Pipe flow visualisations dataset.
\newblock 8 2023.
\newblock \href{https://doi.org/10.15131/shef.data.23791491.v1} {doi:10.15131/shef.data.23791491.v1}.

\end{thebibliography}

\end{document}